\documentclass[11pt,a4paper]{amsart}
\usepackage[top=4cm,bottom=4cm,left=3.5cm,right=3.5cm]{geometry}

\usepackage{subcaption}
\usepackage{tabularx}
\usepackage{amsmath,amssymb,amsthm}
\usepackage[english]{babel}
\usepackage{mathtools}
\usepackage{enumerate}
\usepackage[numbers]{natbib}
\usepackage[hidelinks]{hyperref}
 
\usepackage{comment}
 
\usepackage{multirow}
\usepackage{xcolor}
\usepackage{algorithm}
\usepackage[noend]{algpseudocode}

\newcommand{\F}{\mathbb{F}}

\newcommand{\Z}{\mathbb{Z}}

\newcommand{\mC}{\mathcal{C}}

\newtheorem{theorem}{Theorem}

\newtheorem{lemma}[theorem]{Lemma}
\newtheorem{proposition}[theorem]{Proposition}
\newtheorem{corollary}[theorem]{Corollary}
\theoremstyle{definition}

\newtheorem{definition}[theorem]{Definition}
\theoremstyle{remark}
\newtheorem{remark}[theorem]{Remark}
\newtheorem{example}[theorem]{Example}

\numberwithin{equation}{section}

\newcommand{\zps}{\mathbb{Z} /p^s \mathbb{Z}}
\newcommand{\zpsk}[2][s]{\left(\mathbb{Z}/p^{#1}\mathbb{Z}\right)^{#2}}

\usepackage{stmaryrd}

\def\zs{\left(\mathbb{Z} /{2^s}\mathbb{Z}\right)}

\def\zpsn{\left(\mathbb{Z} /{p^s}\mathbb{Z}\right)^n}

\def\cC{\mathcal{C}}

\def\supp{{\mathrm{supp}}}

\usepackage{longtable}
\usepackage{multirow,multicol}
\usepackage{caption}
\usepackage{ltablex}\keepXColumns

\newcolumntype{Y}{>{\centering\arraybackslash}X}
\newcolumntype{Z}{>{\scriptsize}Y}
\usepackage{rotating}

\usepackage{tikz,pgfplots}
\pgfplotsset{compat=newest}

\title{Bounds in the Lee Metric and Optimal Codes}
\begin{document}
\author[E. Byrne]{Eimear Byrne}
\address{UCD School of Mathematics and Statistics\\
University College of Dublin\\
Ireland
}
\email{ebyrne@ucd.ie}

	\author[V. Weger]{Violetta Weger}
	\address{Department of Electrical and Computer Engineering\\
		Technical University of Munich\\
 Germany
	}
	\email{violetta.weger@tum.de}

\subjclass[2020]{94B05,94B65}

\keywords{ring-linear code, Lee distance, maximum Lee distance, bounds, constant weight codes}

\maketitle

\begin{abstract}
    In this paper we investigate known Singleton-like bounds in the Lee metric and characterize optimal codes, which turn out to be very few. We then focus on Plotkin-like bounds in the Lee metric and present a new bound that extends and refines a previously known, and out-performs it in the case of non-free codes. We then compute the density of optimal codes with regard to the new bound. Finally we fill a gap in the characterization of Lee-equidistant codes.
\end{abstract}

\section{Introduction}\label{sec:intro}

The study of codes that are optimal with respect to some bounds is an important aspect of coding theory. In particular, the study of maximum distance separable (MDS) codes, which are codes that attain the Singleton bound, is a major research area.
In the classical case, where we endow a finite field with the Hamming metric, several properties of MDS codes have been proven, e.g., if $n \leq q $ there exist MDS codes by the construction of \cite{reedsolomon}. It is also well-known that MDS codes are dense as $q$ goes to infinity and that the MDS property is an invariant of duality.
Open questions reamin, such as the famous MDS conjecture \cite{segre}. This conjecture states that MDS codes only exist for $n \leq q+1$ or if $q=2^m$ and $d \in \{4,q-1\}$ and $n=q+2.$ 
This completely changes when we change the metric. In fact, in the rank metric for example we know from \cite{gabidulin, delsarte} that maximum rank distance (MRD) codes exist for any parameters. Moreover, the density of such codes depends on whether they are $\mathbb{F}_{q^m}$-linear or the wider class of $\mathbb{F}_q$-linear rank-metric codes. For the former it was shown in \cite{ale} that MRD codes are dense, while in the latter case we have that MRD codes are neither dense nor sparse \cite{heide2,BR19,heide, anina}. The property on the duals remains true also for this metric, that is the dual of an MRD code is also an MRD code.

One of the questions we want to answer in this paper is how do maximum Lee distance (MLD) codes behave? To answer this question we first have to identify the Singleton-like bounds for the Lee metric for which two proposals are known \cite{shiromoto,alderson}. 
We then show that independently of which bound one considers, MLD codes are sparse. This is done through a characterization of MLD codes, which in certain cases even reduces to only one code and its isomorphic codes (e.g. the original statement of \cite{shiromoto}).
We also provide an answer to the question regarding whether dual codes preserve this property by giving examples of MLD codes whose dual is also MLD and counterexamples. 

Due to these results we then move on to Plotkin-like bounds in the Lee metric. The first such bound was proposed by Wyner and Graham \cite{wyner} and was later improved by Chiang and Wolf \cite{wolf}. However, the Plotkin bound of Chiang and Wolf holds only for free codes. We thus give a generalization of their bound and in addition obtain an improvement, which is Corollary \ref{cor:singK}. We then characterize codes attaining this new bound and compute their density.
For this we use the characterization of Lee-equidistant codes of \cite{wood}. However, this result has a gap: Wood showed that knowing one construction of a shortest length Lee-equidistant code is enough to know all Lee-equidistant codes and characterized these codes for the ambient spaces $\mathbb{F}_p$ and $\mathbb{Z}/2^s\mathbb{Z}$. 
In the general case, i.e., $\mathbb{Z}/p^s\mathbb{Z}$, Wood was able to show that the rank of a linear Lee-equidistant code can be at most 2. We hence provide the two missing constructions of shortest length Lee-equidistant codes, which finally completes the construction and characterization of all Lee-equidistant codes.

This paper is organized as follows: in Section \ref{sec:prelim} we recall the basics of ring-linear coding theory. In Section \ref{sec:init} we then provide some initial observations and techniques of how to obtain the studied bounds, which also aims at putting them in the same context. In Section \ref{sec:overview} we give an overview of the existing Singleton-like and Plotkin-like bounds in the Lee metric and characterize MLD codes. In the main part of this paper, Section \ref{sec:avbds}, we then present the new bound we have obtained, provide the characterization of their optimal codes and compute their density.
Finally, in Section \ref{sec:characterization} we complete the characterization of linear Lee-equidistant codes, which were first examined by Wood \cite{wood}.  

\section{Preliminaries}\label{sec:prelim}
Throughout this paper we will consider codes to be submodules over the integer residue ring $\mathbb{Z}/p^s\mathbb{Z}$, where $p$ is a prime and $s$ is a positive integer. We write $\langle p^i \rangle$ to denote either the ideal $p^i\mathbb{Z}/p^s\mathbb{Z}$ or the submodule $p^i(\mathbb{Z}/p^s\mathbb{Z})^n$, depending on the context.
We denote by $M = \left\lfloor p^s/2 \right\rfloor.$  See \cite{numbermod} for a survey of the topic of linear codes over finite chain rings. A more general treatment of the theory of codes over rings is provided in \cite{nechaev}.

\begin{definition}
A $\mathbb{Z}/p^s\mathbb{Z}$-module  $\mathcal{C}$ of $ \left(\mathbb{Z}/p^s\mathbb{Z}\right)^n$ is called a linear code of length $n$.
\end{definition}
In the classical case over finite fields a code is a linear subspace in $\mathbb{F}_q^n$ and thus has a dimension $k$.
We can take as an analogue of the dimension,  the {\em type} of the code, defined by:
$$k = \log_{p^s}(\mid \mathcal{C} \mid).$$

It is well known that a $\mathbb{Z}/p^s\mathbb{Z}$ module is isomorphic to 
$$\mathcal{C} = \left(\mathbb{Z}/p^s\mathbb{Z}\right)^{k_1} \times \left(\mathbb{Z}/p^{s-1}\mathbb{Z}\right)^{k_2} \times \cdots \times \left(\mathbb{Z}/p\mathbb{Z}\right)^{k_s}.$$
Thus, as an additional parameter of the code we call $(k_1, \ldots, k_s)$ its subtype. It holds that
$$k = \sum_{i=1}^s \frac{s-i+1}{s} k_i.$$
In addition, $k_1$ is called the free rank of the code and $K= \sum_{i=1}^s k_i$ is called its rank.
In \cite{free} the density of free codes was computed, showing that free codes are neither dense nor sparse for $n \to \infty$ and clearly, if $p \to \infty$ free codes are dense.
As in the classical case, a linear code is usually represented through its generator or parity-check matrix.
\begin{definition}
Let $\mathcal{C} \subseteq \left(\mathbb{Z}/p^s\mathbb{Z}\right)^n$ be a linear code. Then we call any matrix $G$ a generator matrix of $\mathcal{C}$, if its row-span is $\mathcal{C}$. In addition, we call any matrix $H$ a parity-check matrix of $\mathcal{C}$, if its null-space is $\mathcal{C},$ equivalently if $GH^\top$ is the zero-matrix. The socle of a linear code is the sum of its minimal submodules; for the codes considered in this paper, the socle of $\mathcal{C}$ is $\langle p^{s-1} \rangle \cap \mathcal{C}$.
\end{definition}
It is often helpful to consider these matrices in their systematic form.

\begin{proposition}
Let $\mathcal{C}$ be a linear code in $\zpsk{n} $ of support subtype $(k_1, \ldots, k_s)$ and rank $K$. Then $ \mathcal{C}$ is permutation equivalent to a code having the following systematic generator matrix $G \in \zpsk{K \times n}$
\begin{equation}\label{systematicformG}
G =\begin{pmatrix}
\text{Id}_{k_1}  & A_{1,2} & A_{1,3} & \cdots & A_{1,s} & A_{1,s+1}  \\
0 & p\text{Id}_{k_2} &  pA_{2,3} & \cdots & pA_{2,s} & pA_{2,s+1}  \\
0 & 0 & p^2 \text{Id}_{k_3} & \cdots & p^2A_{3,s}  & p^2 A_{3,s+1} \\
\vdots & \vdots &  \vdots & & \vdots & \vdots \\
0 & 0 & 0  & \cdots &  p^{s-1} \text{Id}_{k_s} & p^{s-1} A_{s,s+1} 
\end{pmatrix},
\end{equation}
where $A_{i,s+1} \in (\mathbb{Z} / p^{s+1-i}\mathbb{Z})^{k_i \times (n-K)}, A_{i,j} \in (\mathbb{Z} / p^{s+1-i} \mathbb{Z})^{k_i \times k_j}$ for $j \leq s$.
In addition, the code $\mathcal{C}$ is permutation equivalent to a code having the following systematic
 parity-check matrix $H \in \zpsk{(n-k_1) \times n}$,
\begin{equation}\label{systematicformH}
H =\begin{pmatrix}
B_{1,1} & B_{1,2} & \cdots & B_{1,s-1} & B_{1,s} & \text{Id}_{n-K} \\
pB_{2,1} & pB_{2,2} & \cdots & pB_{2,s-1} & p\text{Id}_{k_s} & 0  \\
p^2B_{3,1} & p^2B_{3,2} & \cdots & p^2 \text{Id}_{k_{s-1}} & 0 & 0 \\
\vdots & \vdots & & \vdots & \vdots & \vdots \\
p^{s-1}B_{s,1} & p^{s-1}\text{Id}_{k_2} & \cdots &  0 & 0 & 0
\end{pmatrix},
\end{equation}
where $B_{1,j} \in (\mathbb{Z} / p^s\mathbb{Z})^{(n-K) \times k_j}, B_{i,j} \in (\mathbb{Z} / p^{s+1-i} \mathbb{Z})^{k_{s-i+2} \times k_j}$ for $i > 1$.
\end{proposition}
The dual of the linear code $\mathcal{C}$ is denoted by $\mathcal{C}^\perp$ and defined in the usual way, that is 
$$\mathcal{C}^\perp = \{ x \in \left(\mathbb{Z}/p^s\mathbb{Z}\right)^n \mid x \cdot c = 0  \ \text{for all} \ c \in \mathcal{C}\},$$ where for $x,y \in \left(\mathbb{Z}/p^s\mathbb{Z}\right)^n$ $$x \cdot y = \sum_{i=1}^n x_i y_i.$$
Thus, a parity-check matrix of $\mathcal{C}$ is a generator matrix of $\mathcal{C}^\perp$. 
Furthermore, we have the following parameters for the dual code.

\begin{proposition}
    Let $\mathcal{C} \subseteq \left(\mathbb{Z}/p^s \mathbb{Z}\right)^n$ be a linear code of type $k$, subtype $(k_1,  k_2, \ldots, k_s)$, free rank $k_1$ and rank $K$. Then $\mathcal{C}^\perp$ is a linear code of type $n-k$, subtype $(n-K, k_s, \ldots, k_2)$, free rank $n-K$ and rank $n-k_1.$
\end{proposition}
The ring $\mathbb{Z}/p^s\mathbb{Z}$ can be endowed with many different metrics such as the Hamming metric, the Euclidean metric or the homogeneous metric. In this paper we will focus on the Lee metric. However, this will often be in reference to the Hamming metric. Recall that for $x,y \in \left(\mathbb{Z}/p^s \mathbb{Z}\right)^n$, the Hamming distance between $x$ and $y$ is defined to be
\[
d_H(x,y) = |\{i \in  \{1, \ldots, n \} \ \mid \ x_i \neq y_i|\}.
\]
Furthermore, the Hamming weight of $x$ is $w_H:=d_H(0,x)$. The minimum Hamming distance of $\mathcal{C}$ is $d_H(\mathcal{C}):=\min\{d_H(x,y) \ \mid \ x,y \in \mathcal{C}, x \neq y\}$.

\begin{definition}
   For $x \in \mathbb{Z}/p^s\mathbb{Z}$ we denote by $w_L(x)$ the Lee weight of $x$, which is defined to be:
   $$w_L(x) = \min \{ x, \mid p^s-x\mid \},$$
   where $x$ is interpreted as an integer in $\{0,\ldots,M\}$ in the evalutation 
   $w_L(x)$.
   For $x \in \left(\mathbb{Z}/p^s\mathbb{Z}\right)^n$, the Lee weight is defined additively, that is
   $$w_L(x)= \sum_{i=1}^n w_L(x_i).$$
   The Lee weight induces the Lee distance, i.e., for $x,y \in \left(\mathbb{Z}/p^s\mathbb{Z}\right)^n$ we set
   $$d_L(x,y) = w_L(x-y).$$
\end{definition}
Note that for $x \in \left(\mathbb{Z}/p^s\mathbb{Z}\right)^n$, we have that $$0 \leq w_L(x) \leq nM,$$ and $$w_H(x) \leq w_L(x) \leq Mw_H(x).$$
We endow a linear code with a final parameter: the minimum Lee distance.

\begin{definition}
   Let $\mathcal{C} \subseteq \left(\mathbb{Z}/p^s\mathbb{Z}\right)^n$ be a  linear code. Then the minimum Lee distance of $\mathcal{C}$ is defined as
   $$d_L(\mathcal{C})= \min \{w_L(c) \mid c \in \mathcal{C}, c \neq 0\}.$$
\end{definition}
One can easily observe that 
$$d_H(\mathcal{C}) \leq d_L(\mathcal{C}) \leq M d_H(\mathcal{C}).$$

Recall that a pair of codes are called equivalent with respect to a given metric if there is a linear isometry that preserves the distances between codewords. 
The Extension Theorem holds for both the Hamming and Lee metrics: that is, it is known that any isometry between a pair of codes with respect to either of these metrics extends to one on the whole ambient space \cite{dyshko,macwilliams1962combinatorial}.
For codes with respect to the Hamming metric and isomtery between codes extends to a monomial transformations. A Lee metric isometry is contained in the subgroup of monomial transformations whose non-zero matrix entries are $\pm1$. 
We may also consider isometries between different spaces and for different metrics. One that is relevant here is the Gray map, which is an isometry from $(\Z/4\Z)^n$ endowed with the Lee metric, onto $\left(\mathbb{Z} / 2 \mathbb{Z} \right)^{2n}$ endowed with the Hamming metric.
  
 \section{Initial Observations}\label{sec:init}

 In order to simplify the proofs used in the following bounds we start by making some initial observations and hence unifying the techniques.
For the Hamming metric on $\mathbb{Z}/p^s\mathbb{Z}$ the following Singleton-like bounds are known:
\begin{proposition}\label{MDS}
     Let $\mathcal{C} \subseteq \left( \mathbb{Z}/p^s\mathbb{Z}\right)^n$ be a  code of type $k$, then 
     $$d_H(\mathcal{C}) \leq n-k+1.$$
\end{proposition}
Of course the above bound holds even if the code is not linear. It is well known (see for example \cite{doughertybook, douandshi}) that for linear codes one can also formulate a tighter bound:
\begin{proposition}\label{MDR}
     Let $\mathcal{C} \subseteq \left( \mathbb{Z}/p^s\mathbb{Z}\right)^n$ be a linear code of rank $K$, then 
     $$d_H(\mathcal{C}) \leq n-K+1.$$
\end{proposition}

\begin{proof}
Let $\mathcal{C}'= \mathcal{C} \cap \langle p^{s-1}\rangle.$ Suppose that $\mathcal{C}$ has subtype $(k_1,\ldots, k_s)$ and a generator matrix $G$ in standard form. Then
\[\mathcal{C}' = \left\{ xG \mid x \in p^{s-1}\left(\mathbb{Z}/p^s\mathbb{Z}\right)^{k_1} \times p^{s-2}\left(\mathbb{Z}/p^s\mathbb{Z}\right)^{k_2} \times \cdots \times \left(\mathbb{Z}/p^s\mathbb{Z}\right)^{k_s} \right\},\]
and since 
$|\mathcal{C}'| = p^{k_1+\cdots +k_s} = p^K$ we have that $\mathcal{C'}$ can be identified with an $[n,K]$ linear code over $\F_p$. The result follows by applying the Singleton bound.
\end{proof}
Codes that achieve the bound of Proposition \ref{MDR} are said to have the property of being maximum distance with respect to the rank and are referred to as (MDR) codes, to differentiate them from the usual MDS codes. Note that any linear code that is MDS is also MDR and $\mathcal{C} \subseteq \left(\mathbb{Z}/p^s\mathbb{Z}\right)^n$ is an MDR code if and only if its socle $\mathcal{C}'$ is an MDS code over $\mathbb{F}_p.$ Hence, the same results as in the classical case, i.e., over finite fields apply here. That is, for $n \to \infty$ we have that MDR codes are sparse due to the MDS conjecture \cite{segre} and that they are dense for $p \to \infty.$

Let $\mathcal{C} \subseteq \left(\mathbb{Z}/p^s\mathbb{Z}\right)^n$ be a code. 
Note that whenever we have   
$$d_L(\mathcal{C}) \leq a d_H(\mathcal{C}),$$ for some value $a$, then we can use either the Singleton bound of Proposition \ref{MDS} or the refined Singleton bound for linear codes of Proposition \ref{MDR} for $d_H(\mathcal{C})$ to obtain an immediate upper bound on $d_L(\mathcal{C})$.

\begin{remark}\label{floor}
We remark that since $d_H(\mathcal{C})$ is an integer, for any positive integer $a$, we have $d_H(\mathcal{C}) \geq \left\lceil \frac{d_L(\mathcal{C})}{a} \right\rceil $, whenever
$d_L(\mathcal{C}) \leq a d_H(\mathcal{C})$. This yields 
\[
\left\lfloor \frac{d_L(\mathcal{C})-1}{a} \right\rfloor \leq d_H(\mathcal{C}) -1\] for any such $a$. 
\end{remark}
Regarding the value $a$, one could clearly use $a=M$, being the maximal weight an element of $\mathbb{Z}/p^s\mathbb{Z}$ can achieve. Such arguments will be used to get Singleton-like bounds in the Lee metric. A better bound, however, can be achieved for linear codes using for $a$ the average weight over $\mathbb{Z}/p^s\mathbb{Z}$, thus using a Plotkin-like argument. 
 
For $i \in \{0, \ldots, s-1\}$, let us define $$\widehat{w_L}(i): = \sum\limits_{ a \in \langle p^i \rangle} w_L(a)$$ to be the total weight of the ideal $\langle p^i \rangle.$ An extension of \cite[Theorem 2]{wyner} yields the following result.
\begin{lemma}\label{lem:avideal}
For $i \in \{0, \ldots, s-1\}$ we have that
\[
        \widehat{w_L}(i)
        = \left\{
        \begin{array}{ll}
        \displaystyle
          \frac{p^{2s-i}-p^i}{4}  
          & \text{if } p \text{ is odd},   \\\\
          2^{2s-i-2}    & \text{if } p = 2.
        \end{array}
        \right.
    \]  
    \end{lemma}
Let $\cC$ be a $\zps$-module. 
We define $$\displaystyle \overline{w}_L(\cC):=\frac{1}{|\cC|}\sum_{a \in \cC} w_L(a),$$ that is, $\overline{w}_L(\cC)$ is the average Lee weight of the $\zps$-module $\cC$.   
    
Plotkin-like bounds for linear codes are based on the simple but useful observation that for a linear code $\mathcal{C}$ we have 
\begin{equation*}
    d\frac{|\mathcal{C}|-1}{|\mathcal{C}|} \leq \frac{1}{|\mathcal{C}|} \sum_{c \in \mathcal{C}} w(c) = \overline{w}(\mathcal{C}),
\end{equation*}
where $d$ is the minimum distance of $\mathcal{C}$ with respect to any fixed metric and $w(c)$ is the weight of a word with respect to this metric. 
For the Lee metric, this yields the bound:
\begin{equation}\label{eq:plotkin}
    d_L(\mathcal{C}) \leq \frac{|\mathcal{C}|}{|\mathcal{C}|-1} \overline{w}_L(\mathcal{C}).
\end{equation}
Clearly, this bound is met if and only if every non-zero codeword of $\cC$ has the same Lee weight, i.e., if and only if $\cC$ is equidistant, or in this case Lee-equidistant.

\section{Overview of Existing Bounds}\label{sec:overview}

\subsection{Singleton-like Bounds}

In this section we inspect known Singleton-like bounds for the Lee metric over $\mathbb{Z}/p^s\mathbb{Z}$. 
 
For the most well-known case, i.e., $\mathbb{Z}/4\mathbb{Z}$, the Singleton bound is given through the Gray isometry. 
\begin{theorem}[$\mathbb{Z}/4\mathbb{Z}$-Singleton Bound]\label{Z4SB}
Let $\mathcal{C} \subseteq \left(\mathbb{Z}/4\mathbb{Z}\right)^n$ of type $k$. Then
$$d_L(\mathcal{C}) \leq 2(n-k)+1.$$
\end{theorem}
In fact, let $\mathcal{C}$ be a code in $\left(\mathbb{Z}/4\mathbb{Z}\right)^n$ of type $k= k_1+\frac{1}{2}k_2$, then by applying the Gray isometry we get a code $\mathcal{C}'$  in $\mathbb{F}_2^{2n}$, where it holds that 
$$2^{2k_1+k_2}=\mid \mathcal{C}' \mid \leq 2^{2n-d_H(\mathcal{C})+1} .$$
Since the map is an isometry we have that $d_H(\mathcal{C})=d_L(\mathcal{C})$, thus it holds that 
$$d_L(\mathcal{C}) \leq  2(n-k)+1.$$

\begin{example}
Let us consider the code $\mathcal{C}_1 = \langle (2,2) \rangle$ over $\mathbb{Z}/4\mathbb{Z}$, which has $M=2, n=2, k=1/2$ and $d_L(\mathcal{C})=4$. This code clearly attains the $\mathbb{Z}/4\mathbb{Z}$-Singleton bound of Theorem \ref{Z4SB}.
\end{example}
For rings other than $\mathbb{Z}/4\mathbb{Z}$ there are several possibilities for Singleton-like bounds, such as the bound of Shiromoto \cite{shiromoto}: 
\begin{theorem}\label{shiromoto}
For any code $\mathcal{C}$ in $\zpsk{n}$ of type $k$, we have that 
    $$\left\lfloor \frac{d_L(\mathcal{C})-1}{M}\right\rfloor \leq n-\lceil k \rceil .$$
    \end{theorem}
This bound easily follows from the Singleton bound in the Hamming metric of Proposition \ref{MDS}, the observation that $d_L(\mathcal{C}) \leq Md_H(\mathcal{C})$ and Remark \ref{floor}. Note that in the original statement of \cite{shiromoto}  $k$ is used instead of $\lceil k \rceil.$

\begin{example}
Let us consider the code $\mathcal{C}_2 = \langle (1,2) \rangle$ over $\mathbb{Z}/5\mathbb{Z}$, which has $M=2, n=2, k=1$ and $d_L(\mathcal{C})=3$. This code attains the bound of Theorem \ref{shiromoto}.
\end{example}  
In general, using the Singleton-bound for linear codes of Proposition \ref{MDR} and the fact that $d_L(\mathcal{C}) \leq M d_H(\mathcal{C})$, one immediately gets an improved Singleton-like bound for the Lee metric. 
\begin{corollary}\label{LMDR}
   Let $\mathcal{C} \subseteq \left(\mathbb{Z}/p^s\mathbb{Z}\right)^n$ be a linear code of rank $K$, then 
   $$d_L(\mathcal{C}) \leq M(n-K+1).$$
\end{corollary}
Hence, using Remark \ref{floor} we get that

  \begin{corollary}\label{newSB}
     Let $\mathcal{C} \subseteq \left(\mathbb{Z}/p^s\mathbb{Z}\right)^n$ be a linear code of rank $K$, then 
     $$\left\lfloor \frac{d_L(\mathcal{C})-1}{M} \right\rfloor \leq n-K.$$
 \end{corollary}
 The example $\mathcal{C}_2$ from before also attains this bound. In fact, if  a linear code  $\mathcal{C}$ is such that $d_L(\mathcal{C})=M(n-k+1)$ which is greater or equal to  $M(n-K+1),$ we must have that it attains the bound of Corollary \ref{LMDR}  as well and that $\mathcal{C}$ is a free MDS code.

One can also consider the bound provided by Alderson and Huntemann \cite{alderson}:
\begin{theorem}\label{alderson}
For any code $\mathcal{C}$ in $\zpsk{n}$ of  type $k$ a positive integer, $1<k<n$ we have that $$d_L(\mathcal{C}) \leq M(n-k).$$
\end{theorem}

   \begin{example}
    Let $\mathcal{C}_3 = \langle (2,0,1), (1,3,4)\rangle$ over $\mathbb{Z}/5\mathbb{Z}$. Here we have $n=3, k=2, M=2$ and since $(1,1,0) \in \mathcal{C}_3$ we have $d_L(\mathcal{C})=2.$ This code attains the  bound of Theorem \ref{alderson} since $d_L(\mathcal{C})=2 = M(n-k)=2.$
   \end{example}

\subsection{Characterization}

We now characterize codes attaining the Singleton-like bounds in the Lee metric.
As usual, the notation $[N,S,D]$ is used to denote the parameters of linear code of length $N$, cardinality $q^S$, and minimum distance $D$ for some fixed metric, while the notation $(N,S',D)$ applied for an arbitrary code of length $N$ and cardinality $S'$ and minimum distance $D$.

The linear binary MDS codes have been classified for some time. They are referred to as the trivial MDS codes and have one of the following parameter sets for the Hamming metric: $[N,N,1]$, or $[N,N-1,2]$ or $[N,1,N]$.
Alderson  recently showed in \cite{binMDS} that any binary MDS code is equivalent to a trivial MDS code. Indeed, any binary MDS code has one of the following parameter sets: 
$(N,2^N,1)$, or $(N,2^{N-1},2)$ or $(N,2,N)$.
Due to the nature of the bound of Theorem \ref{Z4SB}, which uses the Gray isometry to identify a $\mathbb{Z}/4\mathbb{Z}$ code of length $N$ for the Lee metric with a binary code of length $2N$ for the  
Hamming metric and since the only MDS codes over the binary are trivial codes, we get the following result.

\begin{proposition}
     The only linear codes that attain the  $\mathbb{Z}/4\mathbb{Z}$-Singleton bound of Theorem \ref{Z4SB} are $\mathcal{C}= \langle(2, \ldots, 2)\rangle$, its dual $\mathcal{C}^\perp$ and the ambient space $\left(\mathbb{Z}/4\mathbb{Z}\right)^{n}$ itself.
\end{proposition}

\begin{proof}
Let $\mathcal{C}$ be an $(n,4^k,d)$ code over $\mathbb{Z}/4\mathbb{Z}$ endowed with the Lee metric. Then $\mathcal{C}$ corresponds to a binary Hamming metric $(2n,2^{2k},d)$ code $\mathcal{C}'$ via the Gray isometry. Therefore, if $\mathcal{C}$ attains the $\mathbb{Z}/4\mathbb{Z}$-Singleton bound then $\mathcal{C}'$ must be a binary MDS code. In the case that the binary code $\mathcal{C}'$ has parameters $(2n,2,2n)$, this  implies that $\mathcal{C}$ has parameters $(n,2,2n)$. Therefore, if $\mathcal{C}$ is linear over $\mathbb{Z}/4\mathbb{Z}$ it must be 
$\langle (2,\ldots,2) \rangle$. In the case where $\mathcal{C}'$ has parameters $(2n,2^{2n-1},2)$, this implies that $\mathcal{C}$ has parameters $(n,4^{n-1/2},2)$ and thus $\mathcal{C}$ corresponds to $\langle (2, \ldots, 2)\rangle^\perp$. Finally, if $\mathcal{C}'$ has parameters $(2n,2^{2n},1)$, we force $\mathcal{C}$ to have parameters $(n, 4^n,1)$, which corresponds to the ambient space $\left( \mathbb{Z}/4\mathbb{Z}\right)^n$.  
\end{proof}
Even though the Singleton-like bound from Theorem \ref{shiromoto} is sharp, there are very few linear codes that attain this bound. We exclude the trivial case $\mathcal{C} = \left(\mathbb{Z}/p^s\mathbb{Z}\right)^n$ of minimum Lee distance 1, which always attains the bound. 

\begin{theorem}\label{charS}
The only linear codes $\mathcal{C} \subset \left(\mathbb{Z}/p^s\mathbb{Z}\right)^n$ of type $k$ and rank $K$ that attain the bound of Theorem \ref{shiromoto} are 
\begin{itemize}
    \item for $p$ odd: codes equivalent to $\mathcal{C}= \langle(1,2) \rangle \subset \left(\mathbb{Z}/5\mathbb{Z}\right)^2$ or over any $p^s$ with $k < \lceil k \rceil =K=n<k+1,$ i.e., $d_L(\mathcal{C})=1,$
    \item for $p=2$: $\mathcal{C}= \langle (2^{s-1}, \ldots, 2^{s-1} ) \rangle$ with $d_L(\mathcal{C})=2^{s-1}n$, or such that $k \neq K = \lceil k \rceil \in \{n,n-1\}$ giving $d_L(\mathcal{C})\leq 2^{s-1}$ and $d_L(\mathcal{C})=2^s$ respectively.
\end{itemize}
\end{theorem}

\begin{proof}
Let us consider a linear code $\mathcal{C} \subset \left(\mathbb{Z}/p^s\mathbb{Z}\right)^n$ of type $k$ and rank $K$. In order for such a code to obtain the bound of Theorem \ref{shiromoto}, we have that $$d_L(\mathcal{C} )= M(n-\lceil k \rceil)+ \alpha,$$ for some $\alpha \in \{1, \ldots, M\}.$
Thus, all codewords have a Hamming weight that is at least $n- \lceil k \rceil +1$ and since $$n-K+1 \geq d_H(\mathcal{C}) \geq n- \lceil k \rceil+1,$$ we get that $\lceil k \rceil =K$ and $\mathcal{C}$ is an MDR code. 

We denote by $c$ a minimal Hamming weight codeword, by $g_i$ the $i$-th row of a generator matrix of $\mathcal{C}$ in systematic form and by $u$ the number entries of a tuple that have Lee weight $M$.

Let us first consider the case where $p$ is odd. Then a minimal Hamming weight codeword $c$ has at least $n-\lceil k \rceil$ positions that are  equal to $\frac{p^s \pm 1}{2}.$ From
$$M(n-\lceil k \rceil)+\alpha \leq w_L(2c) \leq n- \lceil k \rceil + M,$$ we get that $\lceil k \rceil \leq n \leq \lceil k \rceil +1.$

We exclude the case $n=\lceil k \rceil=k,$ which corresponds to the trivial code $\left(\mathbb{Z}/p^s\mathbb{Z}\right)^n.$ The case $n=\lceil k \rceil =K > k $ is however possible and gives $d_L(\mathcal{C})=1$. 

If $n= \lceil k \rceil +1$, then from $$M+ \alpha \leq w_L(2c) \leq 1 +M,$$ it follows that $\alpha=1$ and that $c$ has two non-zero entries, $\frac{p^s \pm 1}{2}$ and $\pm 1$. This implies that
$$M+1 \leq w_L(2c) \leq 1+2,$$ which forces $M\leq 2$ and thus $p^s=5$, for which all codes are free, i.e., $n=k+1$. If $k>1$, however, we could combine two rows of a generator matrix in systematic form to eliminate the entry having Lee weight $M$, that is 
$$M(n-k)+1 \leq w_L(g_i \pm g_j) \leq 2,$$ which is not possible. Thus, we must have $k=1$ and $n=2.$ These codes are all Lee-equivalent to $\mathcal{C}=\langle (1,2) \rangle \subset \left(\mathbb{Z}/5\mathbb{Z}\right)^2.$

In the case $p=2$, we first show that we must have $k \neq K.$
 In fact, if $u<n-k+1$, then
 $$2^{s-1}(n-k)+\alpha \leq w_L(c) \leq u2^{s-1} + (n-k-1-u) (2^{s-1}-1),$$  implies that $n-k \leq n-k-1+\alpha \leq  u-2^{s-1} \leq u$ and
 $$2^{s-1}(n-k)+\alpha \leq w_L(2c) \leq (n-k-1-u)2^{s-1},$$  implies that $2^{s-1}(u+1) \leq \alpha \leq 2^{s-1}$. Thus, $u=0$ and  $ n-k \leq 0,$ which only gives the trivial code $\left(\mathbb{Z}/2^s\mathbb{Z}\right)^n$ of minimum Lee distance 1.
If $u=n-k+1$ and hence $\alpha=2^{s-1}, $
we get from 
$$2^{s-1}(n-k+1) \leq w_L(g_i) \leq 1+u2^{s-1}+(n-k-u)(2^{s-1}-1)$$ that $n-k-1 \leq u-2^{s-1}\leq n-k-2^{s-1}$ that $s=1,$ which we exclude as then we get the Hamming metric.

Hence, it is enough to consider the case $k \neq K.$ Since $\mathcal{C}' = \mathcal{C} \cap \langle 2^{s-1} \rangle$ is a trivial binary MDS code, we must have $K \in \{1,n,n-1\}.$ As a final step of this proof, we will show that if $K=1$ we can only have $\mathcal{C}= \langle (2^{s-1}, \ldots, 2^{s-1})\rangle.$ We clearly have that $K \neq k_1$, since else we would have a free code. We first note that $K=k_s$, since else there would  exist a codeword $\hat{c}$ with $2\hat{c} \neq 0$ and $2 \mid \hat{c}$ and 
$$2^{s-1}(n-1) + \alpha \leq w_L(2\hat{c} ) \leq (n-u)2^{s-1}.$$  This forces $u=0$, but then 
$$2^{s-1}(n-1)+\alpha \leq w_L(\hat{c}) \leq n2^{s-2}$$ forces $n=1$ which we may exclude. 
Finally, since $K=k_s=1$, and $\mathcal{C}=\langle c \rangle$ we note that $c$ has at least $n-1$ positions that are equal to $2^{s-1}$ and one additional non-zero entry $x$. If $x$  was not $2^{s-1}$ we could multiply by $2$ and would get a codeword of Hamming weight 1, which forces $n=1$, as $d_H(\mathcal{C})=n.$
\end{proof}
If we would have kept the original statement of \cite{shiromoto}, that is 
$$d_L(\mathcal{C}) \leq M(n-k)+ \alpha,$$ for some $\alpha \in \{1, \ldots, M\}$, we would get that only linear codes with $p^s=5, n=2$ and $k=K=1,$ i.e., $d_L(\mathcal{C})=3,$  can attain this bound. 
We remark that in \cite[Lemma 13]{alderson}, it was already observed that for $k>1 \in \mathbb{N}$, there is no linear code that attains the bound of Theorem \ref{shiromoto}. We have thus extended their characterization. 
However, also for the bound of Theorem \ref{alderson} from \cite{alderson}, we have that only very few linear codes are optimal:

\begin{theorem}
The only linear codes $\mathcal{C} \subset \left(\mathbb{Z}/p^s\mathbb{Z}\right)^n$ of type $k$ and rank $K$ that attain the bound of Theorem \ref{alderson} are 
\begin{itemize}
    \item for $p$ odd:  codes with $p^s=5, k+1 \leq n \leq k+3$ or free codes with $p^s \in \{7,9\}, n=k+1,$
    \item for $p=2:$  free codes with $s=2, k+1 \leq n \leq k+2$, free codes with $s=3, n=k+1$ or $k+1=K \in \{n,n-1\}.$
\end{itemize}
\end{theorem}

\begin{proof}
Let us consider a linear code $\mathcal{C} \subseteq \left(\mathbb{Z}/p^s\mathbb{Z}\right)^n$ of type $n>k>1 \in \mathbb{N}$ and rank $K$ which attains the  bound of Theorem \ref{alderson}, i.e.,  $$d_L(\mathcal{C} )= M(n-k).$$ 
Again, we have that every codeword has Hamming weight at least $n- k$ and since $$n-K+1 \geq d_H(\mathcal{C}) \geq n-k,$$ we must have that $k \in \{K-1,K\}$. 

We use the same notation as before, that is; we denote by $c$ a minimal Hamming weight codeword, by $g_i$ a row of a generator matrix in systematic form and by  $u$ the number entries of a tuple which have Lee weight $M$.

Let us first consider the case $p$ odd. We first exclude the case $k=K-1$, since then we must have an MDR code and $$ M(n-k) \leq w_L(2c) \leq n-k,$$ which is not possible. 
Thus, we can consider $k=K,$ i.e., we have a free code.  From
$$M(n-k) \leq w_L(g_i) \leq 1+Mu +(n-k-u)(M-1),$$ follows that $u \in \{n-k-1, n-k\}.$ If $u =n-k$, however, we could combine two rows $g_i$ and $g_j$ in such a way that at least one entry cancels, getting that 
$$M(n-k) \leq w_L(g_i \pm g_j) \leq 2 + n-k-1,$$ from which follows that $M=2$ and $n=k+1.$ 

If $u=n-k+1$, then we have to consider an additional entry $x_i$ of Lee weight $<M$. 
From 
$$M(n-k) \leq w_L(g_i) = 1 + (n-k-1)M +w_L(x_i),$$ we get that $w_L(x_i)=M-1$. We again choose $g_i \pm g_j$ in such a way that either $x_i$ intersects with the $n-k$ non-zero positions of $g_j$ and the weight in this intersection becomes 1, or if they do not intersect, we choose $g_i \pm g_j$ such that the entry of weight $M-1$ cancels out, getting that $$M(n-k) \leq w_L(g_i \pm g_j) \leq 2+n-k+1.$$ This implies that $(M-1)(n-k) \leq 3$, which is only possible if $M=2$, i.e., $p^s=5$ and $k+1 \leq n \leq k+3$ or $M\leq 4$ and $n=k+1.$

We can now consider the case  $p=2.$ If $k=K$, i.e., we have a free code, then
$$2^{s-1}(n-k) \leq w_L(g_i) \leq 1+ u2^{s-1} + (n-k-u)(2^{s-1}-1),$$ forces that $u \in \{n-k-1, n-k \}.$ For $u=n-k$ we get that 
$$2^{s-1}(n-k) \leq w_L(2g_i) \leq 2,$$ which forces $s=2$ and $n=k+1.$ 

If $u=n-k+1$, we have an additional entry $x_i$ of Lee weight $<2^{s-1}$ and
$$2^{s-1}(n-k) \leq w_L(g_i) \leq 1+(n-k-1)2^{s-1}+w_L(x_i),$$  shows that $w_L(x_i)=2^{s-1}-1.$ But then we can consider $2g_i$, of Lee weight at most 4, which implies that $s=2$ and $k+1 \leq n \leq k+2$ or $s=3$ and $n=k+1.$

If $k=K-1$, we consider again $\mathcal{C}' = \mathcal{C} \cap \langle 2^{s-1}\rangle$ which is a trivial binary MDS code and hence $K \in \{1,n,n-1\}$. Clearly, we can exclude the case $K=1$ since then $k=0.$
\end{proof}
 In particular, this implies that in the case of $p$ odd, we must have $d_L(\mathcal{C}) \in \{2,3,4,6 \}$ and if $p=2$ we must have $d_L(\mathcal{C}) \in \{2,4, 2^{s-1}, 2^s\}$.
 
 Since an optimal code for the bound of Theorem \ref{shiromoto} is such that $\lceil k \rceil =K$ and 
$d_L(\mathcal{C})= M(n-\lceil k\rceil )+\alpha$ for some $\alpha \in \{1, \ldots, M\}$, 
Theorem \ref{charS} also includes the bound of Corollary \ref{newSB}. 
 \begin{corollary}\label{charrank}
 The only linear codes $\mathcal{C} \subseteq \left(\mathbb{Z}/p^s\mathbb{Z}\right)^n$ that attain the bound of Corollary \ref{newSB} are 
 \begin{itemize}
 \item for $p$ odd: $K=n$ or a code equivalent to $\mathcal{C}= \langle (1,2) \rangle \subset \left(\mathbb{Z}/5\mathbb{Z}\right)^2.$
     \item for $p=2$:  $\mathcal{C}=\langle (2^{s-1}, \ldots, 2^{s-1})\rangle,$ and $K\in \{n,n-1\}.$ 
 \end{itemize}
 \end{corollary}
 We will call a code maximum Lee distance (MLD) if it is optimal code with respect to any of the considered Singleton-like bounds.
 One can immediately see that the density of MLD codes is 0 for $p \to \infty$. For any fixed rate $R=k/n$ or for any fixed $p>7$  we can also see that the density of MLD codes is 0 for $n \to \infty$. 
 
Notice, however, that also with more sophisticated bounds the number of maximum Lee distance codes for $p=2$ will remain small, due to the fact that the socle $ \mathcal{C} \cap \langle 2^{s-1}\rangle$ is a trivial binary MDS code, which cannot be avoided. 
\subsection{Duals of MLD Codes}

The characterizations just shown also answer the question as to whether the dual of MLD codes is also MLD;
for the special case of $\mathbb{Z}/4\mathbb{Z}$, we have seen that non-trivial linear codes that achieve the $\mathbb{Z}/4\mathbb{Z}$-Singleton bound, i.e., $d_L(\mathcal{C}) \leq 2(n-k)+1$, are only  the codes $\mathcal{C}=\langle(2, \ldots, 2)\rangle$ and its dual. Thus, for this particular bound it is true that the dual of an optimal code attains the bound as well, however, for the other Singleton-like bounds, this is (in general) not true, as the choice of   $n$ is very restrictive.

\begin{example}
 Let $\mathcal{C} = \langle (0,1,1), (2,0,0), (0,0,2) \rangle \subset \left(\mathbb{Z}/4\mathbb{Z}\right)^3$. This code attains the bound of Corollary \ref{LMDR} since 
 $ d_L(\mathcal{C})=2 = 2 (n-K+1).$
 However, its dual code $\mathcal{C}^\perp = \langle (2,0,0), (0,2,2)\rangle \subset \left(\mathbb{Z}/4\mathbb{Z}\right)^3$ has $d_L(\mathcal{C}^\perp) = 2$ and thus does not attain the bound since
 $d_L(\mathcal{C}^\perp) =2 < 4 = 2(n-K' +1).$
\end{example}
For a code that attains the bound of Theorem \ref{shiromoto} and $p$ odd, we had two possibilities. In the first one, we note that the code $\mathcal{C}_2= \langle(1,2)\rangle \subset \left(\mathbb{Z}/5\mathbb{Z}\right)^2$ is self-dual, thus the dual attains the bound trivially as well. In the second possibility, we have that $k <\lceil k \rceil =K=n<k+1$. In order to have $n < k'+1 = n-k+1$ as well, we need $k<1$, which forces $n<2$, which we exclude.
If $p=2$, we get again the case $\mathcal{C}= \langle (2^{s-1}, \ldots, 2^{s-1})\rangle$, whose dual has type $n-1/s$ and thus $K'=n$ is satisfied.  
 
 For the bound of Theorem \ref{alderson}, clearly, any code with $n=k+1$ can not have a dual code that attains the bound as well, since the type of the dual code would then be 1, which is excluded.
 Thus, an optimal code with $p$ odd whose dual is also optimal is only possible for $p^s=5$ and in three cases: $n=4$ with $k=2$, or $n=5$ with $k\in\{2,3\}$, or $n=6$ with $k=3.$ 
 In the case  $p=2$, we can only have $n=k+2,$ giving a dual code  of type 2.
 \begin{example}
The code $\mathcal{C}=\langle (1,0,3,4),(0,1,2,3)\rangle \subset \left(\mathbb{Z}/5\mathbb{Z}\right)^4$ attains the bound as $d_L(\mathcal{C})=4= 2(n-k).$
 The dual of $\mathcal{C}$ is given by $\mathcal{C}^\perp = \langle (1,0,2,2), (0,1,4,2) \rangle$ and also attains the bound as $d_L(\mathcal{C}^\perp)=4=2(n-k)$.
  \end{example}

\subsection{Plotkin-like Bounds}
In addition to the Singleton bound we are also interested in the Plotkin bound for the Lee metric and in particular in the techniques used therein.
A first Lee-metric  Plotkin-like bound was provided by Wyner and Graham \cite{wyner}. This bound holds for any code, also non-linear codes and for any type $k.$
For this note that  
$$\overline{w_L}(\mathbb{Z}/p^s\mathbb{Z}) = \begin{cases}  \frac{p^{2s}-1}{4p^s} & \text{if} \ p \ \text{is odd}, \\ 2^{s-2} & \text{if} \ p=2.
 \end{cases} $$
The bound of Wyner and Graham now states that
\begin{theorem}[\cite{wyner}]\label{th:wyner}
For any code $\mathcal{C}$ in $\zpsk{n}$ of type $k$ we have that
$$d_L(\mathcal{C}) \leq \frac{n\overline{w_L}(\mathbb{Z}/p^s\mathbb{Z})}{1- 1/p^{sk}}.$$
\end{theorem}
Finally, we consider the bound of Chiang and Wolf \cite{wolf}, which is an improvement on the Wyner-Graham bound.

\begin{theorem}[\cite{wolf}]\label{CW}
For a free linear code $\mathcal{C}$ in $\zpsk{n}$ of type $k \geq 2$ we have that
 $$\displaystyle d_L(\mathcal{C}) \leq \frac{\overline{w_L}(\mathbb{Z}/p^s\mathbb{Z})(n-k+1)}{1-1/p^s}= \begin{cases}    \displaystyle\frac{p^s+1}{4}(n-k+1) & \text{if} \ p \ \text{is odd}, \\ \\ \displaystyle\frac{2^{2s-2}}{2^s-1}(n-k+1) & \text{if} \ p=2. \end{cases}$$
 \end{theorem}
The difference between Wyner-Graham's bound and Chiang-Wolf's bound is not only that $n$ is replaced by $n-k+1$, but also instead of dividing by $1-1/p^{sk}$ one divides by $1-1/p^{s}$, which is larger whenever $k>1$. The new enumerator stems from the observation that it is enough to consider the Plotkin argument for a subcode $\langle c \rangle.$ 
As observed by Chiang and Wolf, we have that equidistant codes attain this bound. 

Note however, that for this bound Chiang and Wolf have to assume that the code is free.
Indeed we can give an example of a non-free code, where the bound does not hold: let us consider  $\mathcal{C}= \langle (2,2) \rangle$ over $\mathbb{Z}/4\mathbb{Z}.$ In this case we have $k=1/2$, $a=4/3$ and $d_L(\mathcal{C})=4,$ thus we get $d_L(\mathcal{C})=4 \not\leq \frac{4}{3}(2-1/2) =2.$

In what follows, we will generalize the result of Chiang and Wolf to  any (possibly non-free) code.
The version of the Chiang-Wolf bound that works for arbitrary linear codes is as follows.
\begin{proposition}\label{prop:CWk1}
Let $\mathcal{C}$ be a linear code in $\zpsk{n}$ of subtype $(k_1,\ldots,k_s)$, with $k_1 \geq 1$. 
Then
\[
d_L(\mathcal{C}) \leq \begin{cases} \displaystyle \frac{p^s+1}{4}(n-k_1+1) & \text{if} \ p \ \text{is odd}, \\\\                         \displaystyle \frac{2^{2s-2}}{2^s-1}(n-k_1+1) & \text{if} \ p=2. \end{cases}
\]
\end{proposition}
The proof of Proposition \ref{prop:CWk1} is identical to that of the original theorem: choose a parity-check matrix $H$ for the code $\mC$, which has $n-k_1$ rows. Form the $(n-1)\times n$ matrix
$H'$ by appending the rows of the $(k_1 -1)\times n$ matrix $[Id_{k_1-1}\:|\: 0]$ to $H$. The code with parity-check matrix $H'$ is a subcode of $\mC$ and contains a word of Hamming weight at most $n-k_1+1$.

\section{New Bounds}\label{sec:avbds}

If $\cC$ is a $\zps$-submodule of $\zpsn$, then for each ideal $i \in \{0,...,s\}$, we define $$n_i(\cC):=|\{j \in \{1, \ldots, n\}  \ \mid \  \langle \pi_j(\cC)\rangle  = \langle p^i \rangle\}|,$$ where for each $j \in \{1, \ldots, n\}$, $\pi_j$ is the $j$-th coordinate map:
\begin{align*}
    \pi_j: \zpsn &\longrightarrow \zps, \\ (x_1,\ldots,x_n) & \mapsto x_j.
\end{align*}
 If the code $\cC$ is clear from the context, we may simply write $n_i$ in place of $n_i(\cC)$. We say that $\cC$ is non-degenerate if $n_s = 0$.
For a module $\mathcal{C} \subseteq \zpsk{n}$, we call $(n_0, \ldots, n_s)$ its support subtype.
The support of $c \in \zpsk{n}$ is defined to be $\supp(c):=\{ i  \in \{1,...,n\} \mid c_i \neq 0\}$ and the support of the module $\cC$ is $\supp(\cC):=\cup_{c \in \cC} \supp(c)$. That is, the support of $\cC$ is the number of non-zero coordinates of $\cC$. Clearly, $|\supp(\cC)| = n-n_s$.

We will now give an explicit value for $\overline{w_L}(\mathcal{C})$,
generalizing observations made in \cite{wolf}. It shows that for $p=2$, the average Lee weight of a $\zs$-linear code $\mathcal{C}$ is independent of both the subtype $(k_1,\ldots,k_s)$ of the code and on its support subtype $(n_0,n_1,\ldots,n_s)$, but depends only on the support of $\mathcal{C}$. If $p$ is odd, however, this average Lee weight depends on its support subtype.

\begin{lemma}\label{lem:basic}
    Let $\mathcal{C}$ be an $\zps$-submodule of $\zpsn$ of support subtype $(n_0,\ldots,n_s)$. Then 
    \[
        \overline{w}_L(\mathcal{C}) 
        = \sum_{i=0}^{s-1} \widehat{w_L}(i)n_i
        = \left\{
        \begin{array}{ll}
        \displaystyle
          \frac{1}{4p^s} \left(p^{2s}(n-n_s) - \sum_{i=0}^{s-1} p^{2i}n_i\right)  
          & \text{if } p \text{ is odd},   \\\\
          2^{s-2}(n-n_s)
          & \text{if } p = 2.
        \end{array}
        \right.
    \]  
\end{lemma}

\begin{proof}
Recall from Lemma \ref{lem:avideal} that for the ideal $\langle p^i\rangle$, we have:
\[
       \widehat{w_L}(i)
        = \left\{
        \begin{array}{ll}
        \displaystyle
          \frac{p^{2s-i}-p^i}{4}  
          & \text{if } p \text{ is odd},   \\\\
          2^{2s-i-2}    & \text{if } p = 2.
        \end{array}
        \right.
    \]  
    We have $$\overline{w}_L(\mathcal{C}) = |\mathcal{C}|^{-1}\sum_{c \in \mathcal{C}} w_L(c) = |\mathcal{C}|^{-1}\sum_{j=1}^n \sum_{c \in \mathcal{C}} w_L(c_j).$$
    Since $\mathcal{C}$ is linear over $\zps$, for any given $j \in \{1, \ldots, n\}$, we have a module epimorphism 
    $\pi_j:\mathcal{C} \longrightarrow J$ for some ideal $J$ of $\zps$ and hence each element $b$ of $J$ appears
    $$|\{ c \in \mathcal{C} \mid \pi_j(c) = b\}| = |\ker \pi_j \cap \mathcal{C}|$$ times as a coordinate of a codeword of $\mathcal{C}$.
    Therefore, $$\sum_{c \in \mathcal{C}} w_L(c_j) = |\mathcal{C}|/|J| \sum_{a \in J} w_L(a).$$ 
    The result now follows, noting that $|\langle p^i\rangle |= p^{s-i}$.
\end{proof}

\begin{theorem}\label{th:subcode}
    Let $\mathcal{C}$ be a $\zps$-submodule of $\zpsn$ and let $\cC'$ be a non-trivial subcode of $\cC$. Then
    \begin{equation*}
    d_L(\cC) \leq |\cC'|(|\cC'|-1)^{-1} \overline{w}_L(\cC') .
    \end{equation*}
\end{theorem}

\begin{proof}
The claim follows immediately from Lemma \ref{lem:basic} and the fact that $d_L(\cC)\leq d_L(\cC')$.
\end{proof}

\begin{theorem}\label{HammingtoLeei}
    Let $\mathcal{C}$ be a $\zps$-submodule of $\zpsn$. 
    Let $\ell \in \{1,\ldots,s\}$ such that there exists $y \in \mC$ satisfying 
    $w_H(y) = d_H(\langle y \rangle)$ and $y \in \langle p^{s-\ell} \rangle$.
    Then
    \[
       d_L(\mathcal{C}) \leq \left\{
        \begin{array}{ll}
        \displaystyle
           \frac{p^{s-\ell}(p^\ell+1)}{4} d_H(\mathcal{C}) 
          & \text{if } p \text{ is odd},   \\\\
          \displaystyle\frac{2^{s-2+\ell}}{2^\ell-1}d_H(\mathcal{C}) & \text{if } p = 2.
        \end{array}
        \right.
    \]
\end{theorem}

\begin{proof}
   Choose $y \in \mathcal{C}$ such that $w_H(y) = d_H(\mathcal{C})$. 
   Let $\ell \in \{1,\ldots,s\}$ be the least integer such that $p^{\ell} y = 0$. 
   Then $|\langle y \rangle | = p^\ell$ and  
   $y$ has support subtype equal to $(0,\ldots,0,n_{s-\ell},\ldots,n_{s-1},n-d_H(\mathcal{C}))$
   for some $n_j$ satisfying $n_{s-\ell}\geq 1$.
   From  Theorem \ref{th:subcode} and Lemma \ref{lem:basic}, we have
   \[
      d_L(\mathcal{C})
      \leq\left\{
        \begin{array}{ll}
        \displaystyle
          \frac{p^{\ell}}{p^\ell-1} \frac{1}{4p^s}\left( p^{2s}d_H(\mathcal{C}) -p^{2(s-\ell)}d_H(\mathcal{C}) \right)
          & \text{if } p \text{ is odd},   \\\\
         \displaystyle \frac{2^\ell}{2^\ell-1} 2^{s-2}d_H(\mathcal{C})  & \text{if } p = 2.
        \end{array}
        \right.
   \]
   The result now follows by substitution into the previous inequality.
   The case $p=2$ is immediate, while for odd $p$ we have:
   \begin{align*}
       d_L(\mathcal{C})& \leq \frac{p^\ell}{p^\ell-1}\frac{1}{4p^s} d_H(\mathcal{C})(p^s - p^{s-2\ell})
       =  \frac{p^{s-\ell}(p^\ell+1)}{4}d_H(\mathcal{C}).    
   \end{align*}
\end{proof}

For any linear code $\mC \subseteq \zpsn$ it holds that there exist words in $\mC \cap \langle p^{s-1} \rangle$ of Hamming weight equal to $d_H(\mC)$, so that certainly the hypothesis of Theorem \ref{th:subcode} holds with $\ell=1$.
\begin{definition}
   Let $p$ be a prime and let $s$ be a positive integer. Define
   \[
       A(p,s,i):= \left\{
        \begin{array}{ll}
        \displaystyle
           \frac{p^{s-i}(p^i+1)}{4} 
          & \text{if } p \text{ is odd},   \\\\
          \displaystyle\frac{2^{s-2+i}}{2^i-1} & \text{if } p = 2.
        \end{array}
        \right.
    \]
\end{definition}
Note that in respect of this notation, the Chiang-Wolf bound of Proposition \ref{prop:CWk1} is given by:
\begin{align*} 
    d_L(\mC) \leq  \lfloor A(p,s,s) \rfloor (n-k_1+1),
\end{align*}
for a code $\mC$ of subtype $(k_1,\ldots,k_s)$.

Combining Proposition \ref{MDR} and Theorem \ref{th:subcode} we get the following bound. 
\begin{corollary}\label{cor:singK}
     Let $\mathcal{C} \subseteq \left(\mathbb{Z}/p^s\mathbb{Z}\right)^n$ be a linear code of rank $K$.
     Then 
     \begin{align}\label{eqBW}
     d_L(\mathcal{C}) \leq \lfloor A(p,s,1) \rfloor(n-K+1).
     \end{align}
     Let $\ell \in \{1,\ldots,s\}$ such that there exists $y \in \mC$ satisfying 
     $w_H(y) = d_H(\langle y \rangle)$ and $y \in \langle p^{s-\ell} \rangle$. Then 
     \[
       d_L(\mathcal{C}) \leq \lfloor A(p,s,\ell) \rfloor(n-K+1).
     \]
\end{corollary}

Using Remark \ref{floor} we have the following restatement of Corollary \ref{cor:singK}.
\begin{corollary}\label{best}
    Let $\mathcal{C}$ be a $\zps$-submodule of $\zpsn$ of rank $K$. Then
    \[
    \left\lfloor \frac{d_L(\mathcal{C})-1}{A(p,s,1)} \right\rfloor \leq n-K.
    \]
\end{corollary}

In the following figures we compare the new bound of Corollary \ref{cor:singK} to the Singleton-like and Plotkin-like bounds in the Lee metric. For this we fix some ambient space and vary the subtype of the code. We can clearly see that the new bound outperforms all previous bounds for increasing $k_2.$
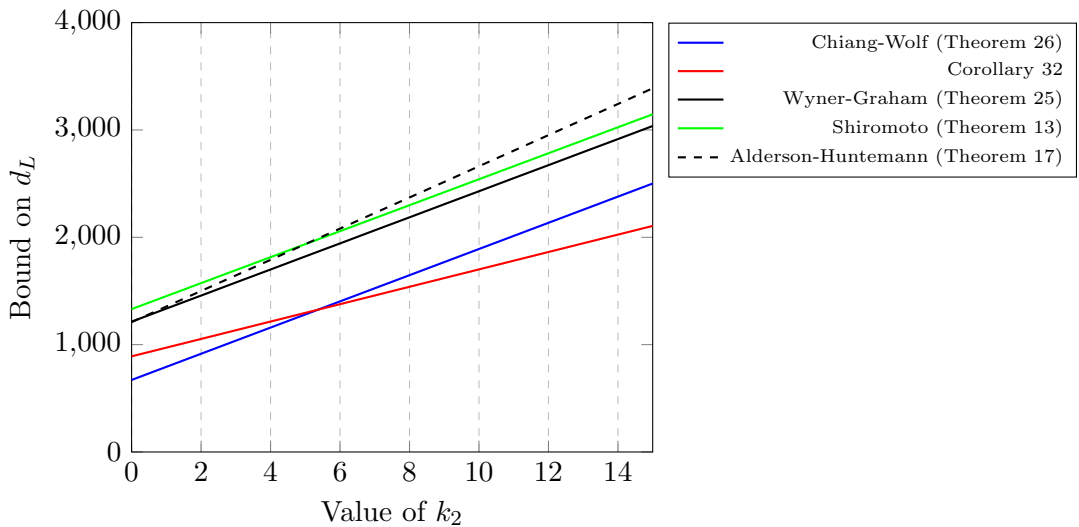
\begin{figure}[h!]
	\centering
	\begin{tikzpicture}[scale=1]
		\begin{axis}[
			legend pos = outer north east,
			legend cell align={right},
			xmin=0, xmax=15,
			ymin=0, ymax=4000,
			xmajorgrids=true,
			grid style=dashed,
			every axis plot/.append style={thick},
			xlabel={Value of $k_2$},
			ylabel={Bound on $d_L$}
			]
			
			\addplot+[color=blue,style = solid,mark=.,mark size=0.5pt]
			coordinates {
				(0 , 671)
                (1 , 793)
                (2 , 915)
                (3 , 1037)
                (4 , 1159)
                (5 , 1281)
                (6 , 1403)
                (7 , 1525)
                (8 , 1647)
                (9 , 1769)
                (10 , 1891)
                (11 , 2013)
                (12 , 2135)
                (13 , 2257)
                (14 , 2379)
                (15 , 2501)	
			};
			\addplot+[color=red,style = solid,mark=.,mark size=0.5pt]
			coordinates {
			    (0 , 891)
                (1 , 972)
                (2 , 1053)
                (3 , 1134)
                (4 , 1215)
                (5 , 1296)
                (6 , 1377)
                (7 , 1458)
                (8 , 1539)
                (9 , 1620)
                (10 , 1701)
                (11 , 1782)
                (12 , 1863)
                (13 , 1944)
                (14 , 2025)
                (15 , 2106)
			};
			\addplot+[color=black,style = solid,mark=.,mark size=0.5pt]
			coordinates {
                (0 , 1214)
                (1 , 1336)
                (2 , 1457)
                (3 , 1579)
                (4 , 1700)
                (5 , 1822)
                (6 , 1943)
                (7 , 2065)
                (8 , 2186)
                (9 , 2308)
                (10, 2429)
                (11, 2551)
                (12, 2672)
                (13, 2794)
                (14, 2915)
                (15, 3037)	
			};
			\addplot+[color=green,style = solid,mark=.,mark size=0.5pt]
			coordinates {
                (0 , 1331 )
                (1 , 1452)
                (2 , 1573)
                (3 , 1694)
                (4 , 1815)
                (5 , 1936)
                (6 , 2057)
                (7 , 2178)
                (8 , 2299)
                (9 , 2420)
                (10, 2541)
                (11, 2662)
                (12, 2783)
                (13, 2904)
                (14, 3025)
                (15, 3146)	
			};
			\addplot+[color=black,style = dashed,mark=.,mark size=0.5pt]
			coordinates {
                (0 , 1210)
                (1 , 1355)
                (2 , 1500)
                (3 , 1645)
                (4 , 1790)
                (5 , 1936)
                (6 , 2081)
                (7 , 2226)
                (8 , 2371)
                (9 , 2516)
                (10, 2662)
                (11, 2807)
                (12, 2952)
                (13, 3097)
                (14, 3242)
                (15, 3388)	
			};
			\legend{\tiny{Chiang-Wolf  (Theorem \ref{CW})},
			\tiny{Corollary \ref{cor:singK}},\tiny{Wyner-Graham (Theorem \ref{th:wyner})},\tiny{Shiromoto (Theorem \ref{shiromoto})},
			\tiny{Alderson-Huntemann (Theorem \ref{alderson})}
			}
		\end{axis}
	\end{tikzpicture}
	\caption{\label{fig:newbds2} 
	Comparison of bounds for codes over $\mathbb{Z}/ {3^5}\mathbb{Z} $ of type $(10,k_2,0,0,0)$ and length $2K,K=10+k_2$.\\
	}
\end{figure}

\begin{figure}[h!]
	\centering
	\begin{tikzpicture}[scale=1]
		\begin{axis}[
			legend pos = outer north east,
			legend cell align={right},
			xmin=0, xmax=15,
			ymin=50, ymax=400,
			xmajorgrids=true,
			grid style=dashed,
			every axis plot/.append style={thick},
			xlabel={Value of $k_2$},
			ylabel={Bound on $d_L$}
			]
			
			\addplot+[color=blue,style = solid,mark=.,mark size=0.5pt]
			coordinates {
				(0 , 104)
                (1 , 117)
                (2 , 130)
                (3 , 143)
                (4 , 156)
                (5 , 169)
                (6 , 182)
                (7 , 195)
                (8 , 208)
                (9 , 221)
                (10, 234)
                (11, 247)
                (12, 260)
                (13, 273)
                (14, 286)
                (15, 299)		
			};
			\addplot+[color=red,style = solid,mark=.,mark size=0.5pt]
			coordinates {
			    (0 , 120)
                (1 , 127)
                (2 , 135)
                (3 , 142)
                (4 , 150)
                (5 , 157)
                (6 , 165)
                (7 , 172)
                (8 , 180)
                (9 , 187)
                (10, 195)
                (11, 202)
                (12, 210)
                (13, 217)
                (14, 225)
                (15, 232)		
			};
			\addplot+[color=black,style = solid,mark=.,mark size=0.5pt]
			coordinates {
                (0 , 187)
                (1 , 199)
                (2 , 212)
                (3 , 224)
                (4 , 237)
                (5 , 249)
                (6 , 262)
                (7 , 274)
                (8 , 287)
                (9 , 299)
                (10, 312)
                (11, 324)
                (12, 336)
                (13, 349)
                (14, 361)
                (15, 374)	
			};
		\addplot+[color=green,style = solid,mark=.,mark size=0.5pt]
			coordinates {
                (0 , 192)
                (1 , 204)
                (2 , 216)
                (3 , 228)
                (4 , 240)
                (5 , 252)
                (6 , 264)
                (7 , 276)
                (8 , 288)
                (9 , 300)
                (10, 312)
                (11, 324)
                (12, 336)
                (13, 348)
                (14, 360)
                (15, 372)		
			};
			\addplot+[color=black,style = dashed,mark=.,mark size=0.5pt]
			coordinates {
                (0 , 180)
                (1 , 198)
                (2 , 216)
                (3 , 234)
                (4 , 252)
                (5 , 270)
                (6 , 288)
                (7 , 306)
                (8 , 324)
                (9 , 342)
                (10, 360)
                (11, 378)
                (12, 396)
                (13, 414)
                (14, 432)
                (15, 450)	
			};
			\legend{\tiny{Chiang-Wolf  (Theorem \ref{CW})},
			\tiny{Corollary \ref{cor:singK}},\tiny{Wyner-Graham (Theorem \ref{th:wyner})},\tiny{Shiromoto (Theorem \ref{shiromoto})},
			\tiny{Alderson-Huntemann (Theorem \ref{alderson})}
			}
		\end{axis}
	\end{tikzpicture}
	\caption{\label{fig:newbds2} 
	Comparison of bounds for codes over $\mathbb{Z}/{5^2}\mathbb{Z}$ of type $(15,k_2)$ and length $2K,K=15+k_2$.\\
	}
\end{figure}
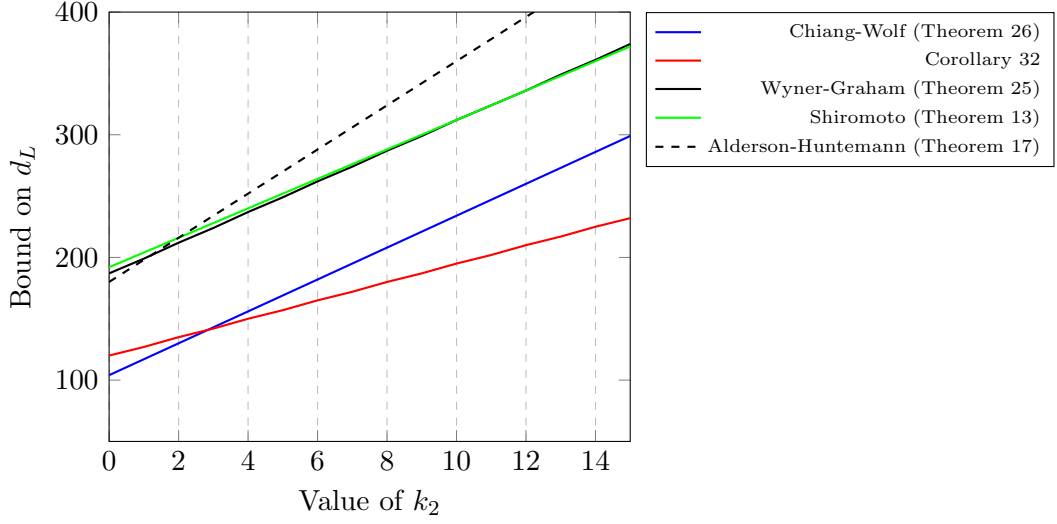

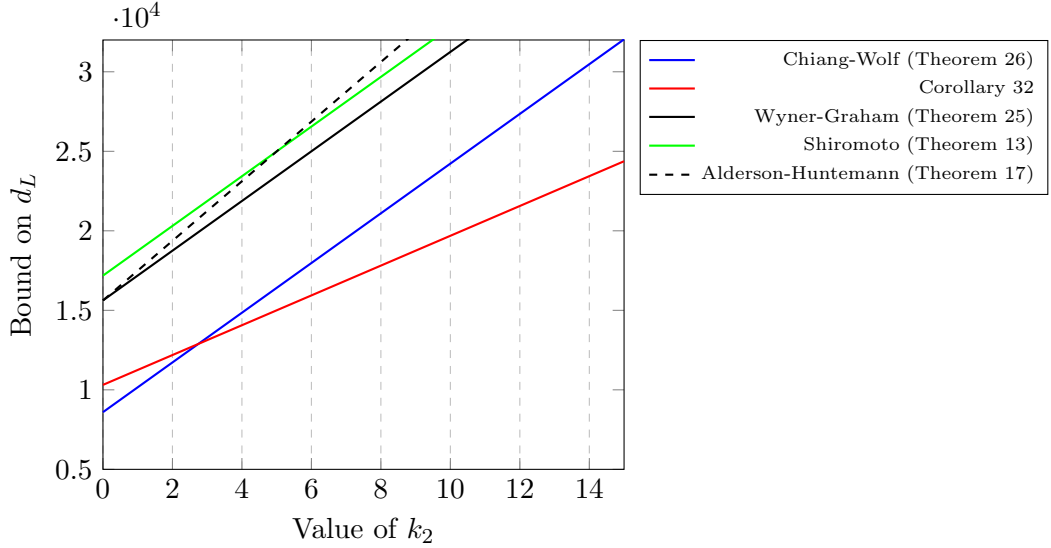
\begin{figure}[h!]
	\centering
	\begin{tikzpicture}[scale=1]
		\begin{axis}[
			legend pos = outer north east,
			legend cell align={right},
			xmin=0, xmax=15,
			ymin=5000, ymax=32000,
			xmajorgrids=true,
			grid style=dashed,
			every axis plot/.append style={thick},
			xlabel={Value of $k_2$},
			ylabel={Bound on $d_L$}
			]
			
			\addplot+[color=blue,style = solid,mark=.,mark size=0.5pt]
			coordinates {
				(0 , 8596)
                (1 , 10159)
                (2 , 11722)
                (3 , 13285)
                (4 , 14848)
                (5 , 16411)
                (6 , 17974)
                (7 , 19537)
                (8 , 21100)
                (9 , 22663)
                (10, 24226)
                (11, 25789)
                (12, 27352)
                (13, 28915)
                (14, 30478)
                (15, 32041)		
			};
			\addplot+[color=red,style = solid,mark=.,mark size=0.5pt]
			coordinates {
			    (0 , 10312)
                (1 , 11250)
                (2 , 12187)
                (3 , 13125)
                (4 , 14062)
                (5 , 15000)
                (6 , 15937)
                (7 , 16875)
                (8 , 17812)
                (9 , 18750)
                (10, 19687)
                (11, 20625)
                (12, 21562)
                (13, 22500)
                (14, 23437)
                (15, 24375)	
			};
			\addplot+[color=black,style = solid,mark=.,mark size=0.5pt]
			coordinates {
                (0 , 15624)
                (1 , 17187)
                (2 , 18749)
                (3 , 20312)
                (4 , 21874)
                (5 , 23437)
                (6 , 24999)
                (7 , 26562)
                (8 , 28124)
                (9 , 29687)
                (10, 31249)
                (11, 32812)
                (12, 34374)
                (13, 35937)
                (14, 37499)
                (15, 39062)	
			};
			\addplot+[color=green,style = solid,mark=.,mark size=0.5pt]
			coordinates {
                (0 , 17182)
                (1 , 18744)
                (2 , 20306)
                (3 , 21868)
                (4 , 23430)
                (5 , 24992)
                (6 , 26554)
                (7 , 28116)
                (8 , 29678)
                (9 , 31240)
                (10, 32802)
                (11, 34364)
                (12, 35926)
                (13, 37488)
                (14, 39050)
                (15, 40612)	
			};
			\addplot+[color=black,style = dashed,mark=.,mark size=0.5pt]
			coordinates {
                (0 , 15620)
                (1 , 17494)
                (2 , 19368)
                (3 , 21243)
                (4 , 23117)
                (5 , 24992)
                (6 , 26866)
                (7 , 28740)
                (8 , 30615)
                (9 , 32489)
                (10, 34364)
                (11, 36238)
                (12, 38112)
                (13, 39987)
                (14, 41861)
                (15, 43736)	
			};
			\legend{\tiny{Chiang-Wolf  (Theorem \ref{CW})},
			\tiny{Corollary \ref{cor:singK}},\tiny{Wyner-Graham (Theorem \ref{th:wyner})},\tiny{Shiromoto (Theorem \ref{shiromoto})},
			\tiny{Alderson-Huntemann (Theorem \ref{alderson})}
			}
		\end{axis}
	\end{tikzpicture}
	\caption{\label{fig:newbds2} 
	Comparison of bounds for codes over $\mathbb{Z}/{5^5}\mathbb{Z}$ of type $(10,k_2,0,0,0)$ and length $2K,K=10+k_2$.\\
	}
\end{figure}
Let us consider two examples, one for $p$ odd and one for $p=2.$
\begin{example}
We consider the Lee-equidistant code $\mathcal{C}=\langle(1,2,1,3)\rangle \subset \left(\mathbb{Z}/5\mathbb{Z}\right)^4$ with minimum Lee distance 6. This code attains both bounds, i.e., that of Corollary \ref{prop:CWk1} and that of Corollary \ref{best}, which are in the case $s=1$ also equal, as 
$$d_L(\mathcal{C})=6 = \frac{6}{4}(4-1+1).$$
\end{example}

\begin{example}
We consider the code  $\mathcal{C} = \langle (0,1,1), (2,0,0), (0,0,2) \rangle \subset \left(\mathbb{Z}/4\mathbb{Z}\right)^3$. This code attains the bound of Corollary \ref{best} for $\ell=1$ since 
 $$ d_L(\mathcal{C})=2 = 2 (n-K+1).$$
 It does not attain the bound of Proposition \ref{prop:CWk1}, as 
 $$d_L(\mathcal{C}) \leq A(4,2,2)(n-k_1+1)=\frac{4}{3}(3-1+1)=4.$$
We also note that we cannot choose $\ell=2$ in Corollary \ref{best}, since the only codewords that have minimal Hamming weight are divisible by 2. 
\end{example}

Using the new bound from Corollary \ref{cor:singK}   we can compare the new bound with the other bounds and the actual maximum Lee distance in Table \ref{table:Z4new}. 
We will denote by SB the classical $\mathbb{Z}/4\mathbb{Z}$-Singleton bound,  by S the bound of Shiromoto in Theorem \ref{shiromoto}, by R-SB the rank-Singleton bound in Theorem \ref{newSB} by AH the bound of Alderson-Huntemann from Theorem \ref{alderson}, by WG the bound of Wyner-Graham of Theorem \ref{th:wyner} and by  CW the  bound by Chiang and Wolf of Theorem \ref{CW}. Maximal $d_L$ denotes the maximal minimum Lee distance obtained by any linear code with the given parameters. We can observe that for these very small parameters, the new bound predicts the actual maximal minimum Lee distance (Column 5) in most of the cases. 
\begin{table}[ht]
 \begin{center}
 \begin{tabular}{|c|c| c|c|c| c| c|c|c|c|c|c|}
 \hline 
  $n$& $K$ & $k$ & $k_1$ & Maximal $d_L$ &  SB &    S & R-SB &  AH & WG &   CW & Corollary \ref{best}\\\hline 
  2 & 1& 1/2 & 0 & 4 &  4&4 &4&   - & 4  & - &  4   \\ 
  2 & 1 &1 &  1&  2 &  3 & 3 &   4  & - & 2  & 2 & 2  \\ 
  2 & 2 & 1 &  0 & 2  & 3  & 3 & 2    &  - &  2 &  -  & 2\\
  2 &  2 &  3/2&  1 &  2 & 2  & 2 &2   & - &  {2}  & - & {2}  \\
  3 & 1 & 1/2 & 0 & 6 & {6}  & {6} &6   & -& 6 &- & {6}   \\
  3 & 1 & 1 & 1 & 4 & 5 & 5 &  6 & - & {4}  & {4} & 4  \\
  3 & 2 & 1 & 0 & 4 & 5 &  5 & {4}  & - & {4}  & - & 4 \\
  3 & 2 & 3/2 & 1 & 2 & 4 & 3 & 3 &- & 3 & - & 3 \\
  3 & 3 & 3/2 & 0 & 2 & 4 & 3 & {2}  & - & 3 & - & 2 \\
  3 & 2 & 2 & 2 & 2   & 3 & 3 & 4 &{2}  & 3 & {2} & 2 \\
  3 & 3 & 2 & 1 & 2   & 3 &3 & {2} & 2  & 3 & - & 2 \\
  3 & 3 & 5/2 & 2 & 2  & {2}  & 2 & 2 & - & 3 & - & {2}  \\
  4 & 1 & 1/2 & 0 & 8 & {8}  & {8} & 8 & - &8 & - & {8}  \\
  4 & 1 & 1& 1 & 5 & 7 & 7 &  8 & - &{5}  & {5} & 5 \\
  4 & 2 & 1 & 0 & 4 & 7 &7 &  5  &- & 5 & - & 5\\
  4 & 2 & 3/2 & 1 & 4 & 6 &5 &  5 & - &4&- & 5 \\
  4 & 3 & 3/2 & 0 & 4 & 6 &5 & {4}  & - &4&- & 4 \\
  4 & 2 & 2 & 2 & 4 & 5 & 5 & 6 & {4}  & {4}  & 4& {4}  \\
  4 & 3 & 2 & 1 & 4 & 5 & 5 & {4}  & 4 & {4}  & - & {4}  \\
  4 & 3 & 5/2 & 2 & 2 & 4 & 3 &  3 & - &4 & - & 3  \\
  4 & 3 & 3 & 3 &2  &3 &  3 & 4  &{2}  & 4 & 2 & {2}  \\
   \hline
  \end{tabular}
\end{center}  
  \caption{Comparison of new bounds for codes over $\mathbb{Z}/4\mathbb{Z}$}
  \label{table:Z4new}
\end{table}

\subsection{Characterization}

    Lee-equidistant codes satisfy (\ref{eq:plotkin}) with equality. One-weight codes have been studied in \cite{wood} under the hypothesis that the MacWilliams Extension Property (MEP) holds. 
    A weight function on $\zpsn$ satisfies the MEP if any $\zps$-linear
    isometry between a pair of codes $\mC$ and $\mC'$ extends to an isometry on $\zpsn$. 
    More recently, in \cite{dyshko} it was shown that the MEP holds for codes over integer modular rings for the Lee metric, answering a conjecture in \cite{wood}. Combining these results gives the following.

\begin{theorem}\label{char} ~
\\
\begin{enumerate}
    \item Let $\mC,\mC'$ be a pair of constant Lee-weight codes of the same subtype. 
    Then $\mC$ and $\mC'$ are equivalent by a Lee-metric monomial transformation. 
    \item If there exists a constant Lee-weight code then there exists non-degenerate linear code $\mC$ of shortest length $n$ of the same subtype that is unique up to equivalence. Any non-degenerate linear constant-Lee-weight code of the same subtype
    is an $\ell$-fold replication of $\mC$, up to equivalence.
\end{enumerate}    
\end{theorem}

\begin{corollary}\label{cor:wood}
     Let $\mC$ be a submodule of $\zpsn$ of rank $K$ and having constant weight. 
     Let $U$ be the collection of orbits of $(\zps)^K$ under the action of $\{1,-1\}$.
     Then one of the following holds.
     \begin{enumerate}
         \item We have $s=1$ and a representative of each member of $U$ appears as a column of a generator matrix for $\mC$ with the same multiplicity.
         \item We have $p=2$ and every column of $\zps^K$ appears with the same multiplicity.
         \item We have $K \leq 2$.
     \end{enumerate}
\end{corollary}

With the characterization of Wood \cite{wood}, we can see that any Lee-equidistant cyclic module over $\mathbb{F}_p$, with $p$ odd, is an $\ell$-fold replication of $\langle(1, \ldots, \frac{p-1}{2})\rangle$ or of an equivalent module. While any Lee-equidistant cyclic module over $\mathbb{Z}/2^s\mathbb{Z}$ is an $\ell$-fold replication of $\langle(1, \ldots, 2^s-1)\rangle$ or of an equivalent module.

We remark that in \cite{equi} Lee-equidistant codes for any rank $K$ were constructed with $n=p^{sK}-1$ for $p$ odd and $s=1$ or for $p=2$. These codes, unfortunately, do not meet the bound of Corollary \ref{cor:singK} since they are not MDR.
The characterization \cite{wood} of Lee-equidistant codes for $K=1$ and $s=1$ is  enough for our purpose, as we only require some $x \in \mathcal{C}' \subseteq \mathbb{F}_p^n$ to generate a Lee-equidistant module. 

An immediate corollary of Theorem \ref{char} and \ref{HammingtoLeei} is given by the following.

\begin{corollary}\label{LeefromHamm}
Let $p$ be an odd prime. Let $x \in \mathbb{F}_p^n$. If $\langle x \rangle$ is Lee-equidistant, then $$w_L(x)=\frac{p+1}{4}w_H(x) .$$
\end{corollary}

Observe that  setting $p=2$ and $\ell=1$ in Corollary \ref{cor:singK} gives the same bound as $p=2$ in Corollary \ref{newSB}, which we have already characterized in Theorem \ref{charrank}. We thus focus on the case $p$ odd.
 
Using Corollary \ref{LeefromHamm} and Theorem \ref{char} we get the following characterization of optimal codes for Corollary \ref{cor:singK}:
\begin{proposition}\label{prop:char}
Let $p$ be an odd prime. Let $\mathcal{C} \subset \left(\mathbb{Z}/p^s\mathbb{Z}\right)^n$ have rank $K$. If $\mathcal{C}$ meets the bound (\ref{eqBW}) of Corollary \ref{cor:singK} then  $n\leq p+1$ and either 
\[
 K=n -p+2 \leq 3\text{ and }d_L(\mathcal{C}) =\frac{p^{s-1}(p^2-1)}{4},
\] or 
\[
K=n+1-\frac{p-1}{2} \leq \frac{p+5}{2} \text{ and }d_L(\mathcal{C})= \frac{p^{s-1}(p^2-1)}{8}.
\]
\end{proposition} 
 
\begin{proof}
Since the code $\mathcal{C}$ achieves the bound \eqref{eqBW}  of Corollary \ref{cor:singK} it must be the case that $\mathcal{C}$ has minimum Hamming distance $n-K+1$. In particular, since words of minimal Hamming weight are contained in $\langle p^{s-1} \rangle \cap \mathcal{C}$, we have that 
$\mathcal{C}$ contains an $[n,K,n-K+1]$ subcode $\mathcal{C}' = \langle p^{s-1} \rangle \cap \mathcal{C}$, which can be identified with an MDS code over $\mathbb{F}_p$. Therefore, $n \leq p+1$ by \cite{ball}.

In addition, in order to attain $d_L(\mathcal{C}) =A(p,s,1)d_H(\mathcal{C})$, it must be the case that there exists $x \in \mathcal{C}'$ of minimal Hamming weight in $\mathcal{C}$ that generates a Lee-equidistant code. Due to the characterization of Theorem \ref{char}, this forces the non-zero entries of $x$ to correspond to repetitions of  $p^{s-1}(\pm 1, \ldots, \pm \frac{p-1}{2})$. Thus, there are only such $x$ if $n \geq \frac{p-1}{2}$. Together with $n \leq p+1$, we have that $x$ must consist of at most two  repetitions of 
$p^{s-1}(\pm 1, \ldots, \pm \frac{p-1}{2})$. It follows that $x$ has Hamming weight  either $p-1$ or $\frac{p-1}{2}$ and so we have $K=n-(p-1)+1 = n-p+2$ or $K=n-(p-1)/2+1 = n-(p-3)/2$. Moreover, we must have 
$w_L(\mathcal{C}) = w_L(x)$ and either $w_L(x)= \frac{p^{s-1}(p^2-1)}{4}= d_L(\mathcal{C}),$ or $w_L(x)= \frac{p^{s-1}(p^2-1)}{8} = d_L(\mathcal{C})$.  
Since $n\leq p+1$, in the first case we get that $K=n-p+1 \leq 3$ and in the second case we have $K= n-(p-1)/2+1 \leq (p+5)/2.$
\end{proof}

We have the following statement on the density of codes that meet the bound of Corollary \ref{cor:singK}.

\begin{theorem}
The density of optimal linear codes $\mathcal{C} \subset \left(\mathbb{Z}/p^s\mathbb{Z}\right)^n$ of rank $K$ for the bound of Corollary \ref{cor:singK} with $\ell=1$ is 0 for   $n$ or $p$ going to infinity.
\end{theorem}

\begin{proof} 
 Since the socle $\mathcal{C}_0$ of an optimal code $\mathcal{C}$ for Corollary \ref{cor:singK} must be an MDS code over $\mathbb{F}_p$ of length $n$, we immediately get that the density of optimal codes is zero as $n \to \infty.$
 Clearly, due to Proposition \ref{prop:char}, if we let $p \to \infty$ but let $n,K$ fixed, we get density 0, as the condition $\frac{p-1}{2} \leq n-K+1 \leq p+1$ would then be violated. Thus, in the following we assume that $n,K$ satisfy the condition of Proposition \ref{prop:char}. From \cite{free} we recall that for $p \to \infty$ the density of free codes is one, thus we will assume that the optimal code is free.  
 
The probability  of having an optimal code, is clearly bounded by the probability of the socle having a codeword of Hamming weight $n-K+1$, which generates a Lee-equidistant code.
 We can consider the socle to be a code over $\mathbb{F}_p$ and the generator matrix   to be in systematic form. The number of codes over $\mathbb{F}_p$ that contain at least one codeword which generates a Lee-equidistant code can be overestimated as the number of $(K-1) \times (n-K)$ matrices times the number of vectors in $\mathbb{F}_p^n$, which generate a Lee-equidistant code.  If we denote this number by $x$, we  observe that $x \leq (n-K)! 2^{n-K}$, since we assumed the systematic form we can only permute  $n-K$ positions and for each element  $a \in \mathbb{F}_p\setminus\{0\}$ we have two choices, i.e., $a$ or $-a$. Using that $n-K+1 \leq p-1$, we get that $x \leq (n-K)! 2^{n-K} \leq (2p)^{n-K-1}$,
 and from this we get the claim, as $$\lim_{p \to \infty} \frac{p^{(n-K)(K-1)}x}{p^{(n-K)K}} \leq \lim_{p \to \infty}  (2p)^{-1} =0.$$
 \end{proof}
 
 \section{Characterization of Lee-equidistant Codes}\label{sec:characterization}

 The characterization of linear Lee-equidistant codes of \cite{wood} is not yet complete. Wood showed that whenever we have the smallest length Lee-equidistant code, then all other Lee-equidistant codes are $\ell$-fold replications of this smallest code (up to equivalence). 
 Due to this result we have that one construction of a smallest length Lee-equidistant code is enough to characterize all of them.
 Wood already gave the smallest length Lee-equidistant codes in the cases $\mathbb{F}_p$ and $\mathbb{Z}/2^s\mathbb{Z}$. 
 We characterize the remaining cases, that is for $p$ odd, $s>1$ and  $K=\{1,2\}$, for the complete the characterization of linear Lee-equidistant codes.

We first treat the case $K=1$, which will also be  helpful for the case $K=2.$ Our approach will be to first find conditions on the support subtype of the code in order to then provide a construction of a minimal-length code. In both cases the main tool is to use Lemma \ref{lem:basic} on a code $\mathcal{C}$ of support subtype $(n_0, \ldots, n_s)$ in the form
\begin{align}\label{eq:crucial}
(\mid \mathcal{C} \mid -1)w= \frac{\mid \mathcal{C} \mid}{4p^s} \sum\limits_{i=0}^{s-1} n_i \left( p^{2s}- p^{2i} \right).
\end{align}

\begin{theorem}\label{thm:K=1}
Let $\mathcal{C} \subseteq \left(\mathbb{Z}/p^s\mathbb{Z}\right)^n$ be a minimal-length linear Lee-equidistant code of rank $K=1$ and minimum Lee distance $w$. Let $i$ be the positive integer such that $k_i =1,$ then $\mathcal{C}$ has support subtype $(0, \ldots, 0, n_{i-1}, \ldots, n_{s-1},0)$ with
\begin{align*}
    w &= \frac{p+1}{4} p^{s-1} n_{i-1}, \\
    n_{i-1} (p-1) &= p^{j-i+2} n_j,
\end{align*}
for all $j \in \{i, \ldots, s-1\}.$
\end{theorem}

\begin{proof}
   We first apply Lemma \ref{lem:basic} to the socle $\mathcal{C}_0= \mathcal{C} \cap \langle p^{s-1} \rangle$, which has size $p$ and support subtype $(0, \ldots, 0, n_{i-1}, z_0)$  and immediately get the first claim
   $$w= \frac{p+1}{4} p^{s-1} n_{i-1}.$$
   To show the second claim we use an inductive argument on $j$. For the base case we  consider the subcode $\mathcal{C}_1 = \mathcal{C} \cap \langle p^{s-2} \rangle$ of size $p^2$ and support subtype $(0, \ldots, 0, n_{i-1}, n_i, z_1)$. Using $w= \frac{p+1}{4}p^{s-1} n_{i-1}$ and applying Lemma \ref{lem:basic} we get that 
   $$n_{i-1}(p-1) = n_ip^2.$$
   Assuming the hypothesis on $j-1$, we now show that the claim also holds for $j$, by applying Lemma \ref{lem:basic} to the subcode $\mathcal{C}_{j-i+1} = \mathcal{C} \cap \langle p^{s-j+i-2}\rangle $ of support subtype $(0, \ldots, 0, n_{i-1}, \ldots, n_j, z_{j-i+1})$ and size $p^{j-i+2}$. We have:

\begin{align*}
    \left(p^{j-i+2}-1\right)w    &= \frac{p^{j-i+2}}{4p^s} \sum\limits_{\ell=i-1}^j n_\ell \left(p^{2s}-p^{2\ell+2s-2j-2} \right), \end{align*} from which follows \begin{align*}
\frac{p+1}{4}p^{s-1}n_{i-1}\left(p^{j-i+2}-1\right) &= \frac{p^{j-i-s+2}}{4} \left(n_{i-1}\left(p^{2s}-p^{2i+2s-2j-4}\right) \right. \\ &  +  \sum\limits_{\ell=i}^{j-1} n_\ell p^{\ell-i+2} \left( p^{2s-\ell+i-2} -p^{\ell+2s-2j+i-4}\right) \\ & \left.  + n_j \left(p^{2s}-p^{2s-2} \right) \right).
\end{align*} We get the claim, as \begin{align*} n_{i-1} \left(p^{j-1+2}-1\right)(p+1)p^{2s-j+i-3}&= n_{i-1}\left(p^{2s}-p^{2i+2s-2j-4}\right)  \\ &   + n_{i-1}(p-1) \sum\limits_{\ell=i}^{j-1}  \left( p^{2s-\ell+i-2} -p^{\ell+2s-2j+i-4}\right) \\ &  +  n_j \left(p^{2s}-p^{2s-2} \right).
   \end{align*}
 
\end{proof}

As we have seen in Theorem \ref{thm:K=1} we have that $p^{s-i+1} \mid n_{i-1}$ and since the socle can be identified with a code over $\mathbb{F}_p$, from Corollary \ref{cor:wood} we have that  $n_{i-1}= \frac{p-1}{4} a,$ for some $a \in \mathbb{N}.$ With this we can exactly determine the support subtype of a smallest-length linear Lee-equidistant code.
\begin{corollary}
    Let $\mathcal{C} \subseteq \left(\mathbb{Z}/p^s\mathbb{Z}\right)^n$ be a minimal-length linear Lee-equidistant code of rank $K=1$ and minimum Lee distance $w$. Let $i$ be the positive integer such that $k_i =1,$ then $\mathcal{C}$ has support subtype $(0, \ldots, 0, n_{i-1}, \ldots, n_{s-1},0)$ with
\begin{align*}
w &=  p^{2s-i} \frac{p^2-1}{8}, \\
    n_{i-1}   &= p^{s-i+1} \frac{p-1}{2}, \\
    n_j &= p^{s-j-1} \frac{(p-1)^2}{2},
\end{align*}
for all $j \in \{i, \ldots, s-1\}.$
\end{corollary}

With this we can finally give the construction of a smallest-length linear Lee-equidistant code of rank 1.
\begin{theorem}\label{constrK1}
    Let $i \in \{1, \ldots, s\}$ and $g \in \langle p^{i-1} \rangle \subseteq \left(\mathbb{Z}/p^s\mathbb{Z}\right)^n$ consist of $p$ repetitions of all elements in $\langle p^{i-1} \rangle \setminus \langle p^i \rangle$ up to $\pm 1$ and $p-1$ repetitions of all elements in $\langle p^j \rangle \setminus \langle p^{j+1} \rangle$ up to $\pm 1$ for all $j \in \{i, \ldots, s-1\}.$ Then $\langle g \rangle$ is a shortest-length linear Lee-equidistant code with $k_i=1$.
\end{theorem}
 
 \begin{proof}
    Let us consider the construction of $g$ where the first $p^{s-i+1}\frac{p-1}{2}$ entries belong to the support subtype $n_{i-1}$, the next $p^{s-i-1}\frac{(p-1)^2}{2}$ entries belong to $n_i$ and so on. Then $g$  has the correct weight $w$, since
    \begin{align*}
        \sum\limits_{j=1}^n w_L(g_j) &= \sum\limits_{j=1}^{n_{i-1}} w_L(g_j) + \sum\limits_{\ell=i}^{s-1}\sum\limits_{j=n_{\ell-1}}^{n_\ell} w_L(g_j) \\ 
        &= \frac{1}{2} \left( p \left( \sum\limits_{a \in \langle p^{i-1} \rangle} w_L(a) -  \sum\limits_{a \in \langle p^{i} \rangle} w_L(a)  \right) \right. \\ & \left.  + (p-1) \sum\limits_{\ell=i}^{s-1} \left(  \sum\limits_{a \in \langle p^{\ell} \rangle} w_L(a) -  \sum\limits_{a \in \langle p^{\ell +1} \rangle} w_L(a)   \right)\right) \\
         &= \frac{1}{8} \left( p^{2s-i+2}-p^{2s-i+1}+p-1  \right. \\ & \left. + (p-1) \sum\limits_{\ell=i}^{s-1} \left( p^{2s-\ell} -p^{2s-\ell-1} +p-1 \right) \right) \\
         &= \frac{1}{8}p^{2s-i}\left(p^2-1\right) = w.
    \end{align*}
    For any $\lambda \in \left( \mathbb{Z}/p^s\mathbb{Z}\right)^\times$, $\lambda g$ will be of the same form, as multiplying by $\lambda$ just results in a permutation. If $\mu=p^\ell \lambda$, for some $\lambda \in \left(\mathbb{Z}/p^s\mathbb{Z}\right)^\times$, then we note that $\mu g= \lambda \left( p^\ell g\right)$ gives the desired form of Theorem \ref{constrK1}   for $k_{i+\ell}=1$.
 \end{proof}

 \begin{example}
 Let us consider $s=2$ and $K=k_1=1$. Then,
 $$x=(1,2,3,4,5,6,7,8,1,2,4) $$
 generates a shortest Lee-equidistant code over $\mathbb{Z}/9\mathbb{Z}$, as we repeat 3 times all units up to $\pm 1$  and  2 times all non-zero non-units up to $\pm 1$.
 \end{example}
 
 \begin{example}
For the case $s=3$ and $K=k_2=1$, we have that
 $$x=(3,3,6,6,9,12,12,15,18,21,24) $$
 generates a Lee-equidistant code over $\mathbb{Z}/27\mathbb{Z}$, as we repeat 3 times all elements in $\langle 3 \rangle \setminus \langle 9 \rangle$ up to $\pm 1 $  and  2 times all elements in $\langle 9 \rangle \setminus \{0\}$ up to $\pm 1$.
 \end{example}

Finally we give the conditions and a construction also for the rank 2 case. First, we show that any linear Lee-equidistant code of rank 2 must have at least one generator in its socle.

\begin{theorem}\label{thm:K=2}
A linear Lee-equidistant code  $\mathcal{C} \subseteq \left(\mathbb{Z}/p^s\mathbb{Z}\right)^n$  of rank $K=2$ with $i\leq j <s $   such that $k_i, k_j \neq 0,$ cannot exist.
\end{theorem}

\begin{proof}
   We are considering $\mathcal{C}= \langle g_1, g_2 \rangle$ with $g_1 \in \langle p^{i-1} \rangle$ and $g_2 \in \langle p^{j-1}\rangle$ for $i \leq j <s$. Then $\mathcal{C}$ is of size $p^{2s-i-j+2}$ and support subtype $(0, \ldots, 0, n_{i-1}, \ldots, n_{s-1},0)$. The subcodes $\langle g_1 \rangle$ and $\langle g_2 \rangle$ both have rank $1$ and have respective support subtypes \begin{align*} (0, \ldots,0, n_{i-1}^{(1)}, \ldots, n_s^{(1)}), \\ (0, \ldots, 0, n_{j-1}^{(2)}, \ldots, n_s^{(2)}). \end{align*}
   By Theorem \ref{thm:K=1} we have, 
   $$p^{j-i+2} n_{j}^{(1)} = n_{i-1}^{(1)} (p-1).$$
   Let us consider the subcode $\mathcal{C}' =p^{s-j}\mathcal{C}= \langle p^{s-j}g_1, p^{s-j}g_2 \rangle$ to find a relation between $n_{i-1}$ and $n_{j-1}.$ The code $\mathcal{C}'$ is of size $p^{j-i+2}$ and support subtype $$(0, \ldots, 0, n_{i-1}, \ldots, n_{j-1}, n_s').$$ Thus, 
   \begin{align*}
       w &= \frac{p+1}{4} p^{s-1}n_{i-1}^{(1)}, \\
       n_\ell &= n_\ell^{(1)} \ \forall \ \ell \in \{i-1, \ldots, j-2\}, \\
       n_{i-1}^{(1)}(p-1) &= p^{\ell-i+2} n_\ell^{(1)} \ \forall \ \ell \in \{i, \ldots, s-1\},
   \end{align*}
     which, by applying Equation \eqref{eq:crucial} leads to $n_{i-1} = n_{j-1}p^{j-i}$.
 
   Finally, we apply Lemma \ref{lem:basic} to the subcode $\tilde{\mathcal{C}} = \langle p^{s-j-1}g_1, p^{s-j-1}g_2\rangle$ of size $p^{j-i+4}$ and support subtype $(0, \ldots, 0, n_{i-1}, \ldots, n_{j}, \tilde{n_s})$, to get that 
   $n_j=0.$
   This, however, leads to a contradiction as $n_j^{(1)} \leq n_j$ and $n_{i-1}(p-1) = p^{j-i+2}n_j^{(1)}.$
\end{proof}

\begin{theorem}
Let $\mathcal{C} = \langle g_1, g_2 \rangle \subseteq \left(\mathbb{Z}/p^s\mathbb{Z}\right)^n$ be a minimal-length linear Lee-equidistant code of rank $K=2$ with minimum Lee distance $w$, with $g_1 \in \langle p^{i-1} \rangle$ and $g_2 \in \langle p^{s-1} \rangle$. Then $\mathcal{C}$ has support subtype $(0, \ldots, 0, n_{i-1}, \ldots, n_{s-1},0)$ and $\langle g_1 \rangle, \langle g_2 \rangle$ have support subtype \begin{align*}
    (0, \ldots, 0, n_{i-1}, \ldots, n_{s-2}, n_{s-1}^{(1)}, n_s^{(1)}), \\
    (0, \ldots, 0, n_{s-1}^{(2)}, n_s^{(2)}), 
\end{align*}
respectively, with 
 \begin{align*}
    w&= p^{s-1} \frac{p+1}{4} n_{i-1}, \\ 
    n_{i-1}&= n_{s-1}^{(2)}, \\
     n_{i-1}  & = n_{s-1}p^{s-i}, \\
     n_{i-1} &= p^{s-i+1} n_s^{(1)}, \\ 
     n_{i-1} &= pn_s^{(2)}, \\ 
          n_{i-1}(p-1) &= p^{\ell-i+2} n_\ell^{(1)}, \\ 
 \end{align*}
 for all $\ell \in \{i, \ldots, s-1\}$. 
\end{theorem}

\begin{proof}
   We first observe that  the claims
   \begin{align*}
   w&= p^{s-1} \frac{p+1}{4} n_{i-1}, \\ 
        n_{i-1}(p-1) &= p^{\ell-i+2} n_\ell^{(1)}, \\ 
   \end{align*}
    for all $\ell \in \{i, \ldots, s-1\}$ follow directly from the case $K=1$ treated in Theorem \ref{thm:K=1}. Since also $w= \frac{p+1}{4}p^{s-1} n_{s-1}^{(2)}$ we also get the claim 
    $$n_{i-1}= n_{s-1}^{(2)}.$$
   To find $n_{s-1}$ we use the relations
   $$  n_{i-1}(p-1) = p^{\ell-i+2} n_\ell, $$
    for all $\ell \in \{i, \ldots, s-2\}$
 and  apply Lemma \ref{lem:basic} to $\mathcal{C}$ to get that
   $$n_{i-1} = n_{s-1}p^{s-i}.$$
   With $n_{s-1}= n_{s-1}^{(1)}+ n_s^{(1)}$ and $n_{s-1}^{(1)}p^{s-i+1}= n_{i-1}(p-1)$ we can also compute
   $$n_{i-1} = p^{s-i+1}n_s^{(1)}.$$ 
   The remaining claim, i.e., $pn_s^{(2)}=n_{i-1}$ finally follows from 
   $$n_s^{(2)}= n_i + \cdots + n_{s-1}.$$ 
\end{proof}

With these conditions on the support subtype we can give the exact support subtype of a shortest-length linear Lee-equidistant code of rank 2. The fact that the socle $\mathcal{C}_0$ can be considered as a code over $\mathbb{F}_p$ tells us that that $n_{i-1}= \frac{p-1}{2}a$, for some $a \in \mathbb{N}$. Moreover, from the relation between $n_{i-1}$ and $n_{s-1}^{(1)}$ in Theorem \ref{thm:K=2} we know that $p^{s-i+1} \mid n_{i-1}.$ We then exploit the relations of in Theorem \ref{thm:K=2} between $n_{i-1}$ with the remaining support subtypes to arrive at the following statement. 

\begin{corollary}
    Let $\mathcal{C} = \langle g_1, g_2 \rangle \subseteq \left(\mathbb{Z}/p^s\mathbb{Z}\right)^n$ be a minimal-length linear Lee-equidistant code of rank $K=2$ and minimum Lee distance $w$, with $g_1 \in \langle p^{i-1} \rangle$ and $g_2 \in \langle p^{s-1} \rangle$. Then $\mathcal{C}$ has support subtype of the form $(0, \ldots, 0, n_{i-1}, \ldots, n_{s-1},0)$ and $\langle g_1 \rangle, \langle g_2 \rangle$ have respective support subtypes
    \begin{align*}
    (0, \ldots, 0, n_{i-1}, \ldots, n_{s-2}, n_{s-1}^{(1)}, n_s^{(1)}), \\
    (0, \ldots, 0, n_{s-1}^{(2)}, n_s^{(2)}), 
\end{align*}
 with 
 \begin{align*}
 w&= p^{2s-i}\frac{p^2-1}{8}, \\
    n_{i-1}&=p^{s-i+1} \frac{p-1}{2}, \\
     n_{s}^{(1)} &= \frac{p-1}{2}, \\ 
     n_{s}^{(2)} &= p^{s-i}\frac{p-1}{2}, \\ 
     n_\ell^{(1)}  & = p^{s-\ell-1}\frac{(p-1)^2}{2}, \\
 \end{align*}
 for all $\ell \in \{i, \ldots, s-1\}$. 
\end{corollary}

We are now ready to give a construction for a shortest-length linear Lee-equidistant code of rank 2.
For this we first need to introduce some notation: let us denote by $U_\ell$ the set of all elements in $\langle p^\ell\rangle \setminus \langle p^{\ell+1
}\rangle$ up to $\pm 1$, which is of size $\alpha_\ell= p^{s-\ell-1}\frac{p-1}{2}$ for all $\ell \in \{i, \ldots, s-1\}.$
For $\ell \in \{i, \ldots, s-1\}$, let us denote by $u_\ell$ the tuple of length $(p-1)\alpha_\ell$ consisting of $p-1$ repetitions of each element in $U_\ell$. Whereas for $\ell=i$, we denote by $\tilde{u}_{i-1}$ the tuple of length $p \alpha_{i-1}$ consisting of $p$ repetitions of all elements in $U_{i-1}.$  Let us denote by $\mathbf{0}$ the tuple consisting of $\frac{p-1}{2}$ zeroes. 
Let us denote by $x$ the length $\frac{p-1}{2}$ tuple consisting of all elements in $U_{s-1}$ and by $y=(x,-x)$ be the tuple consisting of all non-zero elements in the socle and finally by $z=(0,y)$ the tuple consisting of all elements in the socle. 
Let $a$ denote the tuple consisting of $p^{s-i}\frac{p-1}{2}$ repetitions of  $z$, by $b$ the tuple consisting of $\left(p^{s-i}-1\right)\frac{p-1}{2}$ repetitions of $y$ and finally by $c$ the tuple consisting of $\frac{p-1}{2}$ repetitions of $x.$

The construction below, is similar to the simplex code: below the $p$ repetitions of a fixed element in $U_{i-1}$ we have all the elements from the socle and below the $p-1$ repetitions of a fixed element in $U_\ell$ for $\ell \in \{i, \ldots, s-1\}$ we have all the non-zero elements of the socle.

\begin{theorem}\label{constrK2}
The matrix
 $$G = \begin{pmatrix} \tilde{u}_{i-1} & u_i & \cdots & u_{s-1} & \mathbf{0} \\ a &  & b & & c
 \end{pmatrix} $$ generates a Lee-equidistant code over $\mathbb{Z}/p^s\mathbb{Z}$ with $k_i=1,k_s=1.$ 
\end{theorem}

\begin{proof} Let us denote the $i$th row of $G$ by $g_i,$ we define 
$$S_{\ell}:= \{j \in \{1, \ldots, n\} \ \mid \ \langle g_{1j} \rangle = \langle p^{\ell}\rangle \}$$ for $\ell \in \{i-1, \ldots, s-1\}$ and $T:= \supp(g_2).$
    We first want to determine the size of the intersection, i.e., $m = \mid S_{i-1} \cap T \mid$.   For this we consider the socle $\mathcal{C}_0 = \mathcal{C} \cap \langle p^{s-1} \rangle $ of size $p^2$ and support subtype $(0, \ldots, 0, \tilde{n}, \tilde{z})$. Applying Lemma \ref{lem:basic} to $\mathcal{C}_0$ we get that 
 $$w=\tilde{n}\frac{p^s}{4},$$ from which follows that $$ \tilde{n}= n_{i-1}\frac{p+1}{p}.$$ Observe that $\tilde{n} = \mid S_{i-1} \cup T \mid$. 
Since $\tilde{n}= 2n_{i-1}-m$, we get that $m= n_{i-1}\frac{p-1}{p}.$ If $r = \mid T \setminus S_{i-1} \mid$ 
we get that $r= n_{i-1} \frac{1}{p} = n_i + \cdots + n_{s-1}$, thus $n= \tilde{n}$.
With the notation introduced beforehand, we have that  the generator $g_1 \in \langle p^{i-1} \rangle$ needs $p$ copies of each element in $U_{i-1}$ and $p-1$ copies of each element in $U_\ell$, since $n_\ell^{(1)} =p^{s-\ell-1} \frac{(p-1)^2}{2}$ for $\ell \in \{i, \ldots, s-1\}$.

 Clearly, the generators $g_1$ and $g_2$ are of the required form from Theorem \ref{constrK1}. We are left with showing that any linear combination of them also results in a tuple of Lee weight $w.$ Any codeword is of the form $\lambda_1 g_1 + \lambda_2 g_2$ for any $\lambda_1, \lambda_2 \in \mathbb{Z}/p^s\mathbb{Z}$. We observe that if $\lambda_1$ is a unit, 
 then the orbits of its action on the coordinates are the distinct $S_\ell$.
 In particular, $\lambda_1g_1$ has the same form as $g_1$ with respect to $\tilde{u}_{i-1}$, respectively of $u_\ell$ for all $\ell \in \{i, \ldots, s-1\}$.
 Similarly, for $\lambda_2$ a unit, $\lambda_2 g_2$ results in a permutation of the coordinates of $g_2$ again within the described form of $g_2$. Clearly, if $p \mid \lambda_2$ then $\lambda_2 g_2 =0$, and thus this case does not need to be considered. As in the rank 1 case, if $p^\ell \mid \lambda_1$ for some $\ell \in \{1, \ldots, s-1\}$, then $p^\ell g_1 +g_2$ will result  in the same construction as in the statement of the theorem but for $i+\ell$ in place of $i$. Thus, it is enough to consider the codeword $$g_1+g_2=( \tilde{u}_{i-1} +a, (u_i, \ldots, u_{s-1})+b, c).$$
 Note that for all $\ell \in \{i, \ldots, s-2\}$ the addition of elements in the socle will only result in a permutation of $u_\ell$ and also for $\tilde{u}_{i-1}.$ Whereas for $u_{s-1}$ the addition of elements in the socle results in $\frac{p-1}{2}$ zeroes. Thus, the new codeword $g_1+g_2$ is a permutation of $g_1.$
\end{proof}

\begin{example}
In the case $k_1=1=k_2$, we have that the code generated by $$G= \left(\begin{array}{cccccccccccc} 1 & 1 & 1 & 2 & 2 & 2 & 4 & 4 & 4 & 3 & 3 & 0 \\
0 & 3 & 6 & 0 & 3 & 6 & 0 & 3 & 6 & 3 & 6 & 3
\end{array}\right)$$  is Lee-equidistant over $\mathbb{Z}/9\mathbb{Z}$.
\end{example}

\begin{example}
For the case $k_2=2$ over $\mathbb{Z}/9\mathbb{Z}$, the following  matrix generates a Lee-equidistant code
$$G= \begin{pmatrix}  3 & 3& 3 & 3 & 3 & 3 & 0 & 0 \\ 0 & 3 & 6 & 0 & 3 & 6 & 3 & 3  \end{pmatrix}. $$ 
\end{example}

\begin{example}
Let us consider the case $\mathbb{Z}/27\mathbb{Z}$ and $k_1=1, k_3=1$. 
Let $$\tilde{u}_0= (1,1,1,2,2,2,4,4,4, \ldots, 13,13,13), $$ i.e., the tuple consisting of 3 repetitions of all elements in $$ \{1,2,4,5,7,8,10,11,13\},$$ let $u_1=(3,3,6,6,12,12)$ and $u_2=(9,9).$ Let $a$  be the tuple consisting of 9 repetitions of $(0,9,18)$ and $b$ be the tuple consisting of 4 repetitions of $(9,18)$. Then the matrix
  $$ G = \begin{pmatrix}  \tilde{u}_0 & u_1 & & u_2 & 0 \\ a & &b & & 9  \end{pmatrix} $$ generates a Lee-equidistant code. 
\end{example}
\begin{example}
Let us consider the case $\mathbb{Z}/27\mathbb{Z}$ and $k_2=1, k_3=1$. 
 The following matrix  generates a Lee-equidistant code
 $$G=\left(\begin{array}{cccccccccccc}
  3 & 3 & 3 & 6 & 6  & 6 & 12 & 12 & 12 & 9 & 9 & 0 \\ 0 & 9 & 18 & 0 & 9 & 18 & 0 & 9 & 18 & 9 &18  &9
 \end{array}\right).$$
\end{example}

\section*{Acknowledgements}\label{sec:ack} 
The second author  is  supported by the Swiss National Science Foundation grant number 195290.

\bibliographystyle{plain}

\bibliography{References}

\end{document}